\documentclass[final,5p,times,fleqn,twocolumn,numbers]{elsarticle}
\usepackage{amsthm}
\PassOptionsToPackage{intlimits,fleqn}{amsmath}
\usepackage{amsmath,amssymb,mathtools}
\usepackage[T1]{fontenc}
\usepackage[english]{babel}% also gives "Appendix A" a single space (elsarticle adds one)
\allowdisplaybreaks[4]
\newcommand{\m}{\boldsymbol}
\newcommand{\R}{\mathbb{R}}

\usepackage{array,booktabs}
\usepackage{enumitem}
\usepackage[linesnumbered,boxed,commentsnumbered,ruled,vlined,longend]{algorithm2e}
\usepackage{lettrine}
\usepackage{xcolor}
\usepackage{tikz}
\usetikzlibrary{shapes.geometric,arrows.meta,calc,positioning,fit,backgrounds}

\usepackage{caption}
\usepackage{titlesec}
\titlespacing*{\section}{0pt}{*0.3}{*0.2}
\titlespacing*{\subsection}{0pt}{*0.3}{*0.2}
\titlespacing*{\subsubsection}{0pt}{*0.3}{*0.2}

\AtBeginDocument{%
  \setlength{\mathindent}{0pt}%
  \setlength{\abovedisplayskip}{3.7pt}%
  \setlength{\belowdisplayskip}{3.7pt}%
  \setlength{\abovedisplayshortskip}{3.5pt}%
  \setlength{\belowdisplayshortskip}{3.5pt}%
  \setlength{\textfloatsep}{1pt plus 1pt minus 0pt}%     column-wide float to text
  \setlength{\dbltextfloatsep}{6pt plus 2pt minus 2pt}% full-width float to text
  \setlength{\floatsep}{1pt plus 1pt minus 1pt}%         between floats
  \setlength{\intextsep}{1pt plus 1pt minus 1pt}%        around [h] floats
  \setlength{\abovecaptionskip}{2pt}%                    table caption to table
  \setlength{\belowcaptionskip}{1pt}%                    after the caption
}

\definecolor{LCSS}{rgb}{0,0.502,0.675}
\usepackage[colorlinks=true]{hyperref}
\AtBeginDocument{\hypersetup{linkcolor=LCSS, citecolor=LCSS, urlcolor=LCSS, anchorcolor=LCSS, filecolor=LCSS}}
\usepackage[noabbrev]{cleveref}% also places the hyperlink anchors of the theorem environments
\renewcommand{\eqref}[1]{\hyperref[#1]{(\ref*{#1})}}

\newtheorem{myprs}{\textbf{\textcolor{LCSS}{Proposition}}}
\newtheorem{mycor}{\textbf{\textcolor{LCSS}{Corollary}}}
\newtheorem{myrem}{\textbf{\textcolor{LCSS}{Remark}}}
\renewenvironment{proof}{%
  \par\vspace{1pt}%
  \noindent\textbf{\textit{\textcolor{LCSS}{Proof.}}}\enspace%
}{%
  \hfill$\blacksquare$%
  \par\vspace{1pt}%
}
\makeatletter
\def\thm@space@setup{%
  \thm@preskip=1pt plus 1pt minus 1pt%
  \thm@postskip=1pt plus 1pt minus 1pt%
}
\def\@begintheorem#1#2{%
  \item[\hskip\labelsep\textbf{\textcolor{LCSS}{\textit{#1 #2}}}]%
  \@ifnextchar[{\@withinfo}{\textbf{\textcolor{LCSS}{.}}\enspace\ignorespaces}}
\def\@withinfo[#1]{\textbf{\textcolor{LCSS}{\textit{ (#1)}}}\textcolor{LCSS}{.} \ignorespaces}
\makeatother

\newcolumntype{L}{>{\raggedright\arraybackslash}X}

\newcolumntype{C}[1]{>{\centering\arraybackslash}p{#1}}

\newcolumntype{Y}{>{\raggedright\arraybackslash}X}

\newcommand{\dd}{\mathrm{d}}

\newcommand{\col}{\operatorname{col}}

\usepackage{tabularx}
\usepackage{algpseudocode}
\usepackage{placeins}

\journal{Applied Energy}

\begin{document}

\newdimen\origiwspc
\newdimen\origiwstr
\origiwspc=\fontdimen2=\fontdimen3{frontmatter}

\title{\LARGE The Iberian Blackout: A \textit{Black Swan} or a \textit{Gray Rhino}?\tnoteref{t1}\\
A Protection-Aware Dynamic Voltage Security Assessment}

\tnotetext[t1, funding]{A \emph{Black Swan} is a rare, hard-to-predict event with outsized impact that seems obvious only in hindsight. A \emph{Gray Rhino} is a highly probable, high-impact threat that is visible and approaching, yet neglected until it happens.}
\tnotetext[funding]{This work is supported by the National Science Foundation under Grant 2550934.}

\author[inst1]{Abdallah Alalem Albustami}
\affiliation[inst1]{organization={Vanderbilt University},
addressline={Civil and Environmental Engineering Department},
city={Nashville},
postcode={37235},
state={TN},
country={USA}}
\ead{abdallah.b.alalem.albustami@vanderbilt.edu}

\author[inst1]{Ahmad F. Taha}
\ead{ahmad.taha@vanderbilt.edu}

\begin{abstract}
On 28 April 2025, the Iberian power system collapsed after a rapid voltage rise that led to widespread generation disconnections and loss of synchronism. The final report of the ENTSO-E Expert Panel attributes the blackout to several interacting factors including ineffective voltage control, fixed power factor operation, misaligned protection settings, manual shunt reactor switching, and limited observability. These factors raise a general question: given the current state, can the next trip, ramp, topology action, or shunt switching raise the voltages measured by overvoltage relays above their settings before voltage controls respond? We answer it with a protection-aware dynamic voltage security assessment. For each candidate event, the method computes the voltage at every overvoltage relay over the relay time window from a dynamic model of the network, its controls, and its relays. It predicts which relays trip after a first trip, identifies assets whose relay or measurement data are insufficient for a safety conclusion, and, when one exists, computes a minimum-cost fast reactive power action that keeps all relay voltages below their settings. Case studies on four benchmark systems and a 2000-bus system that reproduces the mechanism of the Iberian blackout show that the method flagged every event that caused a secondary trip within the relay window, never declared safe a plant that tripped within that window, and, on the benchmarks, found fast reactive power actions that prevented the trips within the relay window whenever it judged such actions feasible. These results were validated with nonlinear time-domain simulations. The code is publicly available.
\end{abstract}


\end{frontmatter}

\section{Introduction and Paper Contributions}\label{sec:intro}

\lettrine[lraise=0.1,nindent=0em,slope=-.5em]{O}{n} 28 April 2025, at 12:33~CEST, the Iberian power system suffered the most severe European blackout in more than two decades.
According to the final report of the European Network of Transmission System Operators for Electricity (ENTSO-E) Expert Panel, an uncontrolled voltage rise and loss of voltage control, accelerated by generator trips and rapid reductions of their outputs, caused voltage instability and cascading generation disconnections in Spain~\cite[p.~6]{final}. The factual report (October 2025) established the sequence of events~\cite[p.~2]{ENTSOE}, and the final report (March 2026) concludes that the blackout resulted from interactions among voltage control capability, reactive power behavior, protection settings, generator ramps, small embedded generation, topology actions, and observability~\cite[pp.~331--334]{final}. Early studies of the incident include technical assessments~\cite{cetinkaya2025spain}, a conceptual model of an overvoltage-driven blackout with a margin calculation~\cite{rouco2026overvoltage}, and policy and institutional lessons for power system management under rapid decarbonization~\cite{morao2026iberian}.

This paper addresses the broader problem of how operators can detect and prevent a cascade of overvoltage trips before it becomes self-sustaining. Such a cascade usually starts in a high-voltage operating state in which an event (such as a generation trip, a fixed power factor ramp, a shunt switching, a topology change, an exchange reduction, a load or pump disconnection, or a local protection operation) raises the voltage measured by the overvoltage relays of plants or of their collector networks\footnote{This voltage is measured inside the plant or in the network behind the transmission connection point, for example at a wind or photovoltaic (PV) collector bus, on the low-voltage side of a transformer, or at an internal plant bus. It is generally not the transmission bus voltage monitored in the energy management system (EMS).}, while available voltage controls are too slow, saturated, unavailable, or electrically remote. Each trip then removes reactive absorption and raises the voltage at other relays. The operational question is:
\emph{Could an operational tool have detected that the system was becoming vulnerable to overvoltage trips while conventional static voltage checks were still satisfied?}
To answer it, we develop a protection-aware dynamic voltage security assessment. It screens and ranks candidate events and computes mitigating actions, using quantities that operators know or can bound in real time.

\begin{table*}[t]
	\centering
	\caption{Established security assessment methods and the proposed method.}
	\label{tab:comparison}
	\footnotesize
	\setlength{\tabcolsep}{3.5pt}
	\renewcommand{\arraystretch}{1.15}
	\newcolumntype{W}[1]{>{\raggedright\arraybackslash\hsize=#1\hsize\linewidth=\hsize}X}
	\begin{tabularx}{\textwidth}{@{}>{\raggedright\arraybackslash}p{0.105\textwidth}W{1}W{0.9}W{1}W{1.1}>{\raggedright\arraybackslash}p{0.205\textwidth}@{}}
		\toprule
		& \textbf{Static voltage stability}~\cite{ajjarapu1992continuation,gao2002voltage,van2007voltage}
		& \textbf{RTCA}~\cite{stott1987security}
		& \textbf{RMS-based DSA}~\cite{morison2004power}
		& \textbf{Cascading outage analysis}~\cite{vaiman2012risk,bialek2016benchmarking}
		& \textbf{This paper} \\
		\midrule
		Assesses       & loadability margin      & N-1 flows and voltages & voltage, angle, frequency & outage sequence, extent & relay voltage rise vs.\ margin \\
		Relay voltage  & no                      & no                     & if relays modeled         & mainly line relays      & yes, with delays \\
		Dynamics       & no (steady state)       & no (steady state)      & full nonlinear            & quasi-steady state or dynamic~\cite{song2015dynamic,gharebaghi2023approach} & linearized per mode, exact jump at switching events \\
		Switching      & new operating point     & contingency list       & simulated events          & outages                 & events, mode changes \\
		Uncertainty    & stress direction        & deterministic N-1      & scenarios                 & sampled sequences       & data bounds, certificate \\
		Mitigation     & reactive support siting & security-constrained redispatch  & simulated actions         & critical events, risk   & fast reactive actions or slack \\
		Cost per event & continuation power flow & one power flow         & one nonlinear simulation  & many simulations        & sparse linear solves \\
		\bottomrule
	\end{tabularx}
\end{table*}

Established security assessment methods do not answer this question (see Table~\ref{tab:comparison}). Static voltage stability assessment computes the distance to voltage collapse with P-V and Q-V curves, continuation power flow, and modal analysis, and studies long-term load and on-load tap changer (OLTC) dynamics~\cite{kundur,taylor1994power,ajjarapu1992continuation,gao2002voltage,van2007voltage}. It has also been extended to wide-area detection of voltage instability and to converter-driven stability~\cite{kundur2004definition,hatziargyriou2021definition,glavic2009wide}. Real-time contingency analysis (RTCA) checks post-contingency flows and transmission voltages in steady state~\cite{stott1987security}. Neither considers the thresholds of overvoltage relays at collector or internal plant buses. Cascading outage studies follow the redistribution of active power, line overloads, and line protection~\cite{dobson2012estimating,hines2009large,carreras2004complex,vaiman2012risk,bialek2016benchmarking}. In the Iberian event, the cascade propagated differently, where each generation trip removed reactive absorption and raised the voltage at the relays of neighboring plants. Dynamic security assessment (DSA) with root mean square (RMS) simulation~\cite{morison2004power} reproduces such interactions when the models and the sequence of events are known, but it needs a nonlinear simulation for every event, relay setting, and operating point, which limits how many candidate actions can be checked as conditions change.

The proposed method therefore asks two questions for each event: how much of the overvoltage margin of each relay does the event use and can the available controls restore that margin before the relay operates? Three modeling choices follow. First, the monitored voltage is the one measured by the relay, not only the transmission bus voltage. The final report documents disconnections caused by overvoltage on the transformer or downstream side, including cases in which the point-of-connection voltage was unavailable or the protection settings were not aligned with the requirements~\cite[pp.~159--164]{final}. Second, the time horizon matches the time scales of the relays and actuators, since a manually switched shunt reactor and a static synchronous compensator (STATCOM) that responds within hundreds of milliseconds are not equivalent. Third, controller dynamics, control modes, and limiter states are represented when they act within the relay time window, and a controller is held fixed only when it responds more slowly than the relays.

\vspace{0.15cm}
\noindent\textit{\textbf{Relevance beyond Iberia.}}
Overvoltage protection, shunt charging, transformer taps, and reactive limits are not new but the operating environment is changing with more inverter-based resources (IBRs) and distributed generation, fewer synchronous units online in some areas, lighter loading of extra-high-voltage lines, faster active power ramps, more fixed power factor operation, and less visibility of plant-side and distribution-side voltages. Studies of low-inertia systems identify these changes as sources of new dynamic stability problems~\cite{milano2018foundations,tielens2016relevance}, and operating experience with very high instantaneous non-synchronous penetration shows their effect on ancillary service procurement and system stability~\cite{badesa2021ancillary}. These conditions do not make renewable generation the direct cause of the Iberian blackout, but they make high-voltage operation with small overvoltage margins more likely, in states where the limiting voltage may be on the relay side of a transformer and the controls needed to stop the rise may be delayed, saturated, unavailable, or electrically ineffective within the relay window.

\vspace{0.1cm}
\noindent\textit{\textbf{Paper contributions.}} The contributions are as follows.
\begin{itemize}[leftmargin=*]
\item From the ENTSO-E final report, we describe how fixed power factor ramps, loss of reactive absorption, shunt and topology actions, transformer taps, protection settings, voltage controls, and missing observability reduce the overvoltage margins of protected plants, and how relay trips propagate a cascade.

\item We compute the voltage at each relay measurement point over the relay time window, in the discrete operating mode defined by the connected assets, shunts, taps, controller modes, excitation limiters, and protection states.

\item We rank events by their effect on relay margins, predict the propagation of relay trips with a threshold cascade model, treat uncertain or missing relay and measurement data with bounds, and compute fast reactive power actions when the available controls are sufficient.

\item We validate the method against nonlinear time-domain simulation on four benchmark systems and on a 2000-bus system with protected renewable plants, covering cascade prediction, screening of operator actions, uncertain and missing data, and mitigation. On the 2000-bus system, we reproduce the mechanism of the Iberian blackout and show that the method detects the event that starts it. The code is made publicly available~\cite{codes}.
\end{itemize}

\noindent\textit{\textbf{Paper organization.}}
Section~\ref{sec:event} summarizes the event and its root causes. Section~\ref{sec:mechanism} describes the cascade mechanism and the variables that represent it. Section~\ref{sec:model} develops the hybrid differential-algebraic equation (DAE) model and the relay voltage response over the relay time window, and Section~\ref{sec:cascade} the cascade screening and the mitigation problem. Section~\ref{sec:operations} describes the operational use, Section~\ref{sec:case_studies} presents the case studies, Section~\ref{sec:answer} answers the title question, and Section~\ref{sec:conclusion} concludes the paper. \ref{app:systems} describes the test systems.

\section{Iberian Blackout: Chronology and Root Causes}\label{sec:event}

\subsection{Chronology}\label{subsec:chronology}
The incident developed in two stages over about thirty minutes (see Fig.~\ref{fig:timeline}). The first stage was the frequency oscillations and the operators' response to them. The final report describes a forced oscillation of about 0.63~Hz caused by a converter and a later inter-area oscillation of about 0.2~Hz. Operators responded with additional meshing, exchange reduction, shunt reactor switching, and changes of the control of high-voltage direct current (HVDC) links~\cite[pp.~11--12, 19, 69, 75--77]{final}. These actions changed the voltage and reactive power state in which the cascade later happened. Specifically, an action taken to damp oscillations can also reduce overvoltage margins, reactive absorption, and the available voltage control.

In the second stage of the cascade the problem shifted from damping to voltage control. Around 12{:}32 voltages increased quickly, then between 12{:}32{:}00 and 12{:}32{:}48 the output of large renewable plants in Spain operating at fixed power factor decreased by about 500~MW, and distributed wind and solar units changed their operating point rapidly or disconnected in the same interval~\cite[p.~13]{final}. At fixed power factor, the reactive power follows the active power rather than the local voltage so an active power ramp becomes a fast reactive power change while voltage control is already limited.

The first large disconnection by protection occurred shortly after 12{:}32{:}57 in the Granada area. It was injecting 355~MW and absorbing 165~MVAr immediately before the trip, with the 400~kV voltage at 417.9~kV~\cite[p.~14]{final}. The trip removed reactive absorption as well as active generation, and in a high-voltage state the loss of absorption raises the voltage further. More generation disconnections followed about 19~s later, near Badajoz and in other areas~\cite[p.~14]{final}. The system then lost synchronism, the AC interconnections tripped, low frequency demand disconnection (LFDD) and pump disconnection operated, and the Iberian mainland system collapsed.

Fig.~\ref{fig:timeline} summarizes the sequence of events starting from the response to oscillations that changed the operating point to relay trips that produced a fast positive feedback through loss of reactive absorption.

\begin{figure}[t]
\centering
\resizebox{0.8\columnwidth}{!}{%
\begin{tikzpicture}[
	box/.style={rectangle, rounded corners=2pt, fill=#1,
		text width=6cm, minimum height=0.8cm, align=center,
		font=\footnotesize, inner sep=3pt},
	arrow/.style={->, >=latex, thick},
	note/.style={font=\scriptsize\itshape, text=brown!100}
	]
	\node[box=gray!10] (initial) {Normal operations (morning)\\high renewable output, light load, low inertia};
	\node[box=gray!10, below=0.35cm of initial] (osc1) {12{:}03 -- Converter driven forced\\oscillation around 0.63\,Hz};
	\node[box=black!30!orange!10, below=0.35cm of osc1] (action) {Operator response\\meshing, exchange reduction, shunt actions, HVDC mode changes};
	\node[box=black!30!orange!15, below=0.35cm of action] (osc2) {12{:}19 -- Inter area oscillation\\around 0.2\,Hz};
	\node[box=black!30!orange!20, below=0.35cm of osc2] (precascade) {12{:}32:00--12{:}32:57 -- Voltage rise\\fixed PF ramps and embedded generation behavior};
	\node[box=black!30!orange!30, below=0.35cm of precascade] (trigger) {12{:}32:57 -- Granada area overvoltage trip\\355\,MW and 165\,MVAr absorption removed};
	\node[box=black!30!orange!40, below=0.35cm of trigger] (cascade) {12{:}33:16--12{:}33:18 -- Cascading generation trips\\additional reactive absorption removed};
	\node[box=black!30!orange!50, below=0.35cm of cascade] (island) {12{:}33:19--12{:}33:21 -- Loss of synchronism\\interconnections trip, LFDD and pump disconnection act};
	\node[box=black!90, below=0.35cm of island, text=white] (blackout) {\textbf{12{:}33:27 -- Total Blackout}};
	\draw[arrow] (initial) -- (osc1);
	\draw[arrow] (osc1) -- (action);
	\draw[arrow] (action) -- (osc2);
	\draw[arrow] (osc2) -- (precascade);
	\draw[arrow] (precascade) -- (trigger);
	\draw[arrow] (trigger) -- (cascade);
	\draw[arrow] (cascade) -- (island);
	\draw[arrow] (island) -- (blackout);
	\draw[thick, gray!70] ([xshift=-3mm]initial.west) -- ([xshift=-3mm]osc2.west);
	\node[font=\scriptsize\itshape, text=gray!70, rotate=90, anchor=center]
	at ([xshift=-6mm, yshift=-9mm]initial.west |- osc1.center) {\textbf{oscillation response}};
	\draw[thick, brown!100] ([xshift=-3mm]precascade.west) -- ([xshift=-3mm]blackout.west);
	\node[note, rotate=90, anchor=center]
	at ([xshift=-6mm, yshift=-4mm]precascade.west |- cascade.center) {\textbf{voltage/protection cascade}};
	\draw[thick, brown!100] ([xshift=3mm]trigger.east) -- ([xshift=3mm]blackout.east);
	\node[note, rotate=270, anchor=center]
	at ([xshift=6mm, yshift=-6mm]cascade.north east) {\textbf{$\sim$30\,s collapse}};
\end{tikzpicture}}
\caption{Sequence of events. The values in the blocks of the final minute are from the management summary and event descriptions of the ENTSO-E final report~\cite[pp.~13--15, 128]{final}, and the oscillation and operator response blocks follow its analysis of the oscillations and mitigation actions~\cite[pp.~11--12, 69, 75--77]{final}.}
\label{fig:timeline}
\end{figure}

\subsection{Root causes relevant to the screening problem}
\label{subsec:rootcause_summary}
The root cause tree of the final report breaks its summary statement~\cite[p.~6]{final} into interacting factors in voltage control, reactive power behavior, topology and oscillation response, protection settings, local generation networks, system defence, embedded generation, and observability~\cite[pp.~331--334]{final}. Its operational, organizational, and procedural findings, such as coordination between transmission and distribution system operators (TSOs and DSOs) and data exchange practices, are outside the scope of this paper. The technical findings fall into five groups that the method proposed herein must represent.

\vspace{0.15cm}
\noindent\textit{\textbf{Oscillation response and operating trajectory.}}
The actions taken against the oscillations changed the electrical state in which the final voltage rise developed. In the method, they appear as changes of topology, operating point, control mode, and candidate disturbances.

\vspace{0.15cm}
\noindent\textit{\textbf{Voltage control and reactive capability.}}
The final report identifies insufficient voltage control from conventional generators, fixed power factor operation of renewable plants, manual shunt reactor switching that required decision and processing time, and the absence of ramping limits for fixed power factor generators~\cite[pp.~331--334]{final}. It also states that manual voltage control was too slow for the final 38~s, and that on this time scale only the dynamic voltage control of generators, STATCOMs, and HVDC links could have limited the voltage deviations~\cite[p.~328]{final}. This group determines which devices can affect the voltage within the relay time window.

\vspace{0.15cm}
\noindent\textit{\textbf{Reactive balance and topology.}}
In the final minute, fixed power factor ramps, loss of reactive absorption after generation trips, charging of energized extra-high-voltage circuits, lower series reactive losses at lower line loading, shunt reactor status, and load or pump disconnection raised the voltage. This group defines the candidate disturbances.

\vspace{0.15cm}
\noindent\textit{\textbf{Relay margins and local networks.}}
The final report documents disconnections caused by overvoltage protection on the transformer, evacuation grid, or downstream side, and lists the voltage control design of local generation networks, overvoltage protection settings, disconnection of small embedded generation, and the small margin between operating and disconnection voltages as root causes~\cite[pp.~146,151,159--160,170--173,331--334]{final}. This group defines the relay voltages, thresholds, delays, and margins.

\vspace{0.15cm}
\noindent\textit{\textbf{Frequency response, distributed energy resources, and observability.}}
Pump disconnection and LFDD operated as part of the Spanish System Defence Plan. LFDD operated correctly at the transmission level, but overvoltages were widespread before and during the load disconnection~\cite[pp.~316--318]{final}. The report also notes limited DSO oscillograph records, reverse power blocking, and limited visibility of distributed energy resources (DER) below 1~MW~\cite[pp.~317--318,463--464]{final}. This group determines which load and embedded generation disconnections are disturbances and which missing data enter the uncertainty set.

\begin{figure*}[t]
	\centering
	\resizebox{0.88\textwidth}{!}{%
		\begin{tikzpicture}[
			>=Stealth, font=\footnotesize,
			node distance=12mm and 12mm,
			block/.style={rectangle,draw=none,fill=white,align=center,
				minimum height=10mm,minimum width=25mm},
			longblock/.style={rectangle,draw=none,fill=black!7,align=center,
				minimum height=10mm,minimum width=25mm},
			ampb/.style={rectangle,draw=none,fill=orange!95!black!15,align=center,
				minimum height=10mm,minimum width=25mm},
			scb/.style={rectangle,draw=none,fill=purple!15,align=center,
				minimum height=10mm,minimum width=25mm},
			ctrlblock/.style={rectangle,draw=none,fill=teal!20,align=center,
				minimum height=10mm,minimum width=25mm},
			mainblock/.style={rectangle,draw=none,fill=red!25,align=center,minimum height=10mm,minimum width=25mm},
			sum/.style={circle,draw=none,fill=teal!100,minimum size=7mm,inner sep=0pt,font=\footnotesize,text=white},
			sig/.style={-Latex,black,thick},
			meas/.style={-Latex,black!70!black,thick,dashed},
			lab/.style={font=\small,inner sep=1pt},
			legendlab/.style={font=\fontsize{8pt}{11pt}\selectfont,inner sep=1pt},
			tago/.style={circle,fill=orange!90!black!40,inner sep=0.7pt,
				font=\fontsize{7pt}{13pt}\selectfont,minimum size=1.75mm},
			tags/.style={circle,fill=purple!35,inner sep=0.7pt,
				font=\fontsize{7pt}{13pt}\selectfont,minimum size=1.75mm}
			]
			\node[lab] (Vref) at (-10mm,0) {$V_{\mathrm{ref}}$};
			\node[sum,right=9mm of Vref] (Se) {$\sum$};
			\draw[sig] (Vref) -- node[lab,above,pos=0.15,xshift=6mm,yshift=1mm] {$+$} (Se);
			\node[ctrlblock,right=7mm of Se] (C) {Voltage Control\\(AVR/STATCOM/SVC)};
			\draw[sig] (Se) -- node[lab,above,xshift=-1mm,yshift=1mm] {$e$} (C);
			\node[sum,right=25mm of C] (Sq) {$\sum$};
			\draw[sig] (C) -- node[lab,above,pos=0.2,xshift=18.5mm,yshift=-4mm] {$+$}
			node[lab,font=\scriptsize, below,xshift=-4mm,yshift=-1.5mm] {\textcolor{purple!55}{finite speed}} (Sq) node[lab,below,xshift=-10mm,yshift=4mm] {$Q_{\mathrm{cmd}}$} (Sq);
			\node[longblock,right=10mm of Sq] (G) {Transmission\\Network};
			\draw[sig] (Sq) -- node[lab,below,pos=0.5,xshift=-1mm,yshift=4mm] {$Q_{\mathrm{net}}$} (G);
			\node[lab,right=16mm of G] (Vt) {$V_t$};
			\draw[sig] (G) -- (Vt);
			\draw[meas] ([xshift=5mm]G.east) |- ++(0,-7mm) -| node[lab,below,pos=0.95,xshift=-3mm,yshift=1mm] {$-$} (Se.south);
			\node[mainblock,below=6.5mm of G, draw=black, thick, label={[black,font=\scriptsize, right =14.5mm,yshift=-2.5mm]\textcolor{purple!55}{unobserved}}] (H) {Collector / downstream\\Overvoltage};
			\node[mainblock,left=5mm of H] (P) {Protection\\Trigger};
			\node[mainblock,left=5mm of P] (T) {Generation\\Trips};
			\draw[sig] (G.south) -- (H.north) node[lab,font=\scriptsize,below,pos=0.55,xshift=8mm,yshift=0.75mm] {fixed taps};
			\draw[sig] (H) -- (P);
			\draw[sig] (P) -- (T);
			\draw[sig] (T.north) |- ++(0,3mm) -| node[lab,above,pos=0.25,xshift=21.5mm,yshift=2.5mm] {$+$}
			node[lab,below,pos=0.55,xshift=6mm,yshift=0.2mm] {$\Delta Q_{\mathrm{trip}}$} (Sq.south);
			\coordinate (s0) at ($(Sq.north)+(-95mm,5mm)$);
			\coordinate (s1) at ($(Sq.north)+(-67mm,5mm)$);
			\coordinate (s2) at ($(Sq.north)+(-39mm,5mm)$);
			\coordinate (s3) at ($(Sq.north)+(-11mm,5mm)$);
			\coordinate (s4) at ($(Sq.north)+(17mm,5mm)$);
			\coordinate (s5) at ($(Sq.north)+(45mm,5mm)$);
			\node[scb,above=7mm of s0] (brp) {Fixed PF Ramps\\Small PV Trips};
			\node[scb,above=7mm of s1] (ufls) {Load / Pump\\Shedding};
			\node[ampb,above=7mm of s2] (mesh) {Network\\Meshing};
			\node[ampb,above=7mm of s3] (export) {Exchange\\Reduction};
			\node[ampb,above=7mm of s4] (react) {Shunt\\Reactors};
			\node[ampb,above=7mm of s5] (hvdc) {HVDC\\Control Mode};
			\draw[thick] (brp.south) -- ++(0,-6mm) -| (s0)
			node[lab,left,pos=0.05,xshift=-2pt] {$-$}
			node[lab,above,pos=0.85,xshift=-20pt,yshift=8pt] {$\Delta Q_{\text{abs}}\!\downarrow$};
			\draw[thick] (ufls.south) -- ++(0,-6mm) -| (s1)
			node[lab,left,pos=0.05,xshift=-2pt] {$-$}
			node[lab,above,pos=0.85,xshift=-20pt,yshift=8pt] {$\Delta Q_{\text{load}}\!\downarrow$};
			\draw[thick] (mesh.south) -- ++(0,-6mm) -| (s2)
			node[lab,left,pos=0.05,xshift=-2pt] {$+$}
			node[lab,above,pos=0.85,xshift=-20pt,yshift=8pt] {$\Delta Q_{\text{cap}}\!\uparrow$};
			\draw[thick] (export.south) -- ++(0,-6mm) -| (s3)
			node[lab,left,pos=0.05,xshift=-2pt] {$-$}
			node[lab,above,pos=0.85,xshift=-20pt,yshift=8pt] {$\Delta Q_{\text{line}}\!\downarrow$};
			\draw[thick] (react.south) -- ++(0,-6mm) -| (s4)
			node[lab,left,pos=0.02,xshift=-2pt,yshift=2pt] {$\pm$}
			node[lab,above,pos=0.85,xshift=-20pt,yshift=8pt] {$Q_{\text{shunt}}$};
			\draw[sig] (hvdc.south) -- ++(0,-6mm) -| (G.north)
			node[lab,font=\scriptsize,above,pos=0.75,xshift=24pt,yshift=8pt]{mode dependent};
			\draw[thick] (s0) -- (s4);
			\draw[sig] ($(Sq.north)+(0,5mm)$) -- (Sq.north);
			\node[tago] at ($(mesh.north east)+(-1mm,-1mm)$)  {{{OA1}}};
			\node[tago] at ($(export.north east)+(-1mm,-1mm)$) {{{OA2}}};
			\node[tago] at ($(react.north east)+(-1mm,-1mm)$)  {{{OA3}}};
			\node[tago] at ($(hvdc.north east)+(-1mm,-1mm)$)   {{{OA4}}};
			\node[tags] at ($(brp.north east)+(-1mm,-1mm)$)    {{{SC2}}};
			\node[tags] at ($(C.south east)+(18mm,2mm)$)    {{{SC1}}};
			\node[tags] at ($(H.south east)+(7mm,3mm)$)    {{{SC3}}};
			\coordinate (leg) at ([xshift=0mm,yshift=12mm]current bounding box.south west);
			\node[mainblock,minimum width=3mm,minimum height=3mm,anchor=south west] at (leg) {};
			\node[legendlab,anchor=west] at ([xshift=4mm,yshift=1.5mm]leg) {Positive feedback};
			\node[ampb,minimum width=3mm,minimum height=3mm,anchor=south west] at ([yshift=-3.8mm]leg) {};
			\node[legendlab,anchor=west] at ([xshift=4mm,yshift=-2.3mm]leg) {Operator actions};
			\node[scb,minimum width=3mm,minimum height=3mm,anchor=south west] at ([yshift=-7.6mm]leg) {};
			\node[legendlab,anchor=west] at ([xshift=4mm,yshift=-6.1mm]leg) {Structural conditions};
			\node[ctrlblock,minimum width=3mm,minimum height=3mm,anchor=south west] at ([yshift=-11.4mm]leg) {};
			\node[legendlab,anchor=west] at ([xshift=4mm,yshift=-9.9mm]leg) {Negative feedback};
	\end{tikzpicture}}
	\caption{Voltage control and protection as feedback loops. OA1--OA4 are operator actions, and SC1--SC3 are structural conditions that weaken the negative feedback or expose downstream protection. Relay trips create the positive feedback, since each trip removes reactive absorption and raises the voltage at other relays.}
	\label{fig:feedback}
\end{figure*}

\section{Overvoltage Mechanisms and Modeling Variables}\label{sec:mechanism}
% =====================================================================
The rest of the paper addresses \emph{overvoltage cascades}, which are sequences of overvoltage relay trips in which each trip removes reactive absorption and raises the voltage at other relays. Such cascades propagate through the loss of reactive absorption and they develop within relay time windows that are short enough for fast controls to limit the voltage rise but too short for slow ones. The binding voltage is the one measured by the relay, on the collector, downstream, or transformer low-voltage side, not the transmission voltage. The Iberian blackout is one instance of this mechanism. Thermal overloads, classical P-V voltage collapse, frequency-only events, and cascades of line distance protection are outside the scope of this work. In particular, this section expresses the five groups of Section~\ref{subsec:rootcause_summary} as variables that can be computed from an operating point, namely the voltage raising changes of the reactive power balance, the voltage margin at the relay measurement point, and the controls that can lower the voltage within the relay time window.

\subsection{Reactive power balance and topology}\label{subsec:reactive_balance}

We use $\Delta q>0$ for a voltage-raising disturbance, which can be an increase of net capacitive injection, a reduction of inductive absorption, or a change of network loading that reduces series reactive losses. These mechanisms are physically different, but they have a similar effect in an area whose voltage is already high.

\vspace{0.15cm}
\noindent\textit{\textbf{Fixed power factor ramps.}} For a unit operating at a constant power factor (PF), define
\(\kappa_j:=\tan\!\big(\arccos(|\mathrm{pf}_j|)\big)\). Then
\begin{equation}\label{eq:fixed_pf}
Q_j=\sigma_j\kappa_j P_j,\qquad \Delta Q_j=\sigma_j\kappa_j\Delta P_j,
\end{equation}
where \(\sigma_j=+1\) denotes reactive injection and \(\sigma_j=-1\) reactive absorption in the power flow sign convention. In this convention, \(\Delta Q_j>0\) raises the voltage, so an output reduction of a unit that absorbs at fixed power factor (\(\sigma_j=-1\)) is a voltage-raising disturbance with \(\Delta q=\Delta Q_j>0\). At \(|\mathrm{pf}|=0.98\), \(\kappa_j\approx0.203\), so the reduction of about 500~MW of fixed power factor output between 12{:}32{:}00 and 12{:}32{:}48 (Section~\ref{subsec:chronology}) changed the reactive power by about 100~MVAr~\cite[p.~13]{final}.

\vspace{0.15cm}
\noindent\textit{\textbf{Generation and collector trips.}} If a tripping element absorbed $Q_j^{\mathrm{abs}}>0$ immediately before disconnection, the voltage-raising part of the trip is
\begin{equation}\label{eq:trip_q}
\Delta q_j=Q_j^{\mathrm{abs}} .
\end{equation}
This term drives the positive feedback, in which each overvoltage trip removes absorption, the voltage rises further, and the margins of neighboring relays shrink. The Granada trip, with 165~MVAr of absorption~\cite[p.~14]{final}, is an example.

\vspace{0.15cm}
\noindent\textit{\textbf{Network meshing, line charging, and exchange reduction.}} A transmission line affects the voltage through its shunt charging and its loading-dependent series reactive losses. For a lossless \(\pi\)-model line \(\ell=(a,b)\) with series reactance \(X_\ell>0\) and total charging susceptance \(B_{c\ell}>0\), let \(Q_{\ell,a}^{\mathrm{net}}\) be the reactive power absorbed by the line from bus \(a\). Then
\begin{equation}\label{eq:line_q_split}
Q_{\ell,a}^{\mathrm{net}}=
\underbrace{\frac{V_a^2-V_aV_b\cos(\theta_a-\theta_b)}{X_\ell}}_{\text{series terminal contribution}}
-
\underbrace{\frac{B_{c\ell}}{2}V_a^2}_{\text{shunt charging injection}} ,
\end{equation}
where the first term is the series contribution at terminal \(a\) and the second is the charging injected at that terminal. The series contributions of both terminals add up to the nonnegative series reactive loss
\begin{equation}\label{eq:line_series_total}
Q_{\ell}^{\mathrm{ser}}=
\frac{V_a^2+V_b^2-2V_aV_b\cos(\theta_a-\theta_b)}{X_\ell}
=X_\ell |I_{\ell}^{\mathrm{ser}}|^2 .
\end{equation}
Energizing an extra-high-voltage line adds the charging term and changes the series coupling in the admittance matrix. At light load, charging can exceed the series loss, and the line then acts as a net reactive source. The final report states that connecting lines decreases the system impedance and increases damping but also increases reactive power production and voltage~\cite[p.~32]{final}. An exchange reduction lowers the branch currents and hence the series loss \(X_\ell |I_{\ell}^{\mathrm{ser}}|^2\), while the charging stays the same, so its net effect on the voltage depends on the operating point. The method represents line energization as a change of admittance and mode, and exchange reduction as a disturbance profile around the current operating point.

\vspace{0.15cm}
\noindent\textit{\textbf{Shunt reactors.}} A shunt reactor with susceptance magnitude $B_{\mathrm{sh}}>0$ absorbs approximately
\begin{equation}\label{eq:reactor}
Q_{\mathrm{abs}}^{\mathrm{sh}}(V)=B_{\mathrm{sh}}V^2 .
\end{equation}
Disconnecting it removes this absorption, a voltage-raising disturbance $\Delta q\approx B_{\mathrm{sh}}(V^0)^2$, and reconnecting it has the opposite effect. Reactor switching also changes the voltage sensitivity of the network through \(\partial Q_{\mathrm{abs}}^{\mathrm{sh}}/\partial V=2B_{\mathrm{sh}}V\), so shunt status enters the method both as a disturbance and as a change of the network mode. Because manual reactor switching takes time~\cite[pp.~331--334]{final}, a reactor is a disturbance when its switching has already happened and a control only on horizons longer than its actuation delay.

\vspace{0.15cm}
\noindent\textit{\textbf{Low frequency demand and pump disconnection.}} In the method, LFDD and pump disconnection are load disturbances, not frequency events. A disconnected block with active and reactive components \(\Delta P_L(t)\) and \(\Delta Q_L(t)\) is represented by
\begin{equation}\label{eq:load_shed_channel}
\Delta \m d_L(t)=\col\big(\Delta P_L(t),\Delta Q_L(t)\big),
\end{equation}
and its effect on the relay voltages is computed with the model of Section~\ref{sec:model}. Disconnections of small embedded generation are represented in the same way, with their pre-event injections.

Fig.~\ref{fig:feedback} shows these effects as feedback loops. OA1--OA4 are operator actions (network meshing, exchange reduction, shunt reactor status, and HVDC control mode), and SC1--SC3 are structural conditions (slow or ineffective voltage control, the reactive behavior of fixed power factor plants and small PV, and the exposure of collector or downstream protection). Voltage control is a negative feedback loop. The cascade is a positive feedback loop, because each relay trip removes reactive absorption. The question is whether the negative feedback available within the relay window dominates the positive feedback.

\subsection{Relay voltage margins, transformer taps, and observability}\label{subsec:protection_horizons}

The overvoltage margin is defined at the voltage measured by the relay and not at a transmission bus.

\vspace{0.15cm}
\noindent\textit{\textbf{Relay voltage and overvoltage margin.}} Let $z_i$ be the magnitude of the voltage measured by the overvoltage relay of asset $i$, called the \emph{relay voltage}, which is the voltage of a collector bus, a transformer low-voltage side, an internal plant bus, a 220~kV evacuation grid, or a distribution interface. The overvoltage margin is
\begin{equation}\label{eq:guard_general}
h_i=V_i^{\mathrm{trip}}-z_i^0 .
\end{equation}
The relay is below its pickup if $h_i>0$, and the smaller $h_i$, the smaller the disturbance that makes it operate.

\vspace{0.15cm}
\noindent\textit{\textbf{Transformer taps.}} If the protected side is behind a step-down transformer with an off-load tap or an OLTC with ratio $n_i$, then approximately
\begin{equation}\label{eq:tap_relation}
z_i=V_{c,i}=\frac{V_{t_i}}{n_i},
\end{equation}
where $V_{t_i}$ is the transmission-side voltage. Equation~\eqref{eq:tap_relation} only expresses that the relay measurement point matters, and the transformer model enters through the algebraic equations of Section~\ref{sec:model}. OLTCs are slow by design, since each tap step is preceded by an intentional delay (typically 30--120~s) and followed by a mechanical switching time. Over a relay window of less than a second to a few seconds, the tap ratio is therefore constant, and the relay voltage follows the transmission voltage through a fixed $n_i$. If the tap was set for an earlier, lower voltage, $n_i$ is too small for the present voltage, so the relay voltage starts closer to its threshold. A transmission-side rise $\Delta V_{t_i}$ then raises it by $\Delta V_{t_i}/n_i$, which exceeds $\Delta V_{t_i}$ when $n_i<1$. The relay voltage can then cross its threshold while the transmission voltage is still within the range accepted by a static check~\cite[pp.~159--164]{final}.

\vspace{0.15cm}
\noindent\textit{\textbf{Measured and reconstructed relay voltages.}} Relay voltages are often not measured in the EMS. The final report describes incomplete or heterogeneous data from DSOs, generators, and other parties, and the need to obtain additional records during the investigation~\cite[pp.~7--8]{final}. When the relay voltage is not measured, the method reconstructs it as
\begin{equation}\label{eq:hidden_output_envelope}
z_i=\psi_i(\m y;\theta_i^{\mathrm{obs}}),\qquad
\theta_i^{\mathrm{obs}}\in\Theta_i^{\mathrm{obs}},
\end{equation}
where \(\theta_i^{\mathrm{obs}}\) contains the unknown measurement point, tap position, transformer ratio, and reconstruction error. For a fixed tap,
\begin{equation}\label{eq:hidden_tap_envelope}
z_i=\frac{V_{t_i}}{n_i}+\varepsilon_i,\qquad
n_i\in[\underline n_i,\overline n_i],\quad
\varepsilon_i\in[\underline\varepsilon_i,\overline\varepsilon_i],
\end{equation}
where \(n_i\) is the turns ratio between the transmission side and the protected side. Here $\varepsilon_i$ must also cover the voltage drop across the transformer, which can be several hundredths of a per unit. If the active and reactive power $P_i+jQ_i$ delivered by the plant at the transmission side is measured and the transformer series impedance $Z_i$, referred to the protected side, is known, the drop can be computed instead. With $V_{t_i}/n_i$ as phase reference, the plant current is $(P_i-jQ_i)n_i/V_{t_i}$, and
\begin{equation}\label{eq:pq_reconstruction}
z_i=\left|\frac{V_{t_i}}{n_i}+Z_i\,\frac{(P_i-jQ_i)\,n_i}{V_{t_i}}\right| ,
\end{equation}
whose error follows from the errors of the measurements, the tap, and $Z_i$. Relay data are bounded in the same way,
\begin{equation}\label{eq:protect_interval_intro}
V_i^{\mathrm{trip}}\in[\underline V_i^{\mathrm{trip}},\overline V_i^{\mathrm{trip}}],\qquad
T_i^{\mathrm{delay}}\in[\underline T_i,\overline T_i].
\end{equation}
The worst-case margin at the operating point is
\begin{equation}\label{eq:worst_case_margin}
\underline h_i=
\inf_{\theta_i\in\Theta_i}
\left(V_i^{\mathrm{trip}}(\theta_i)-z_i^0(\theta_i)\right),
\end{equation}
where \(\theta_i\) collects the reconstruction and relay setting uncertainties. If \(\underline h_i\le0\), the data do not guarantee a positive margin at the operating point. The asset is then classified as having \emph{insufficient data}, and the method cannot certify that it will not trip.

\begin{figure}[t]
\centering
\resizebox{\columnwidth}{!}{%
\begin{tikzpicture}[
>=Stealth,
font=\footnotesize,
bus/.style={
    circle,draw=black!55,fill=white,thick,
    minimum size=10mm,inner sep=0pt
},
line/.style={black!70,thick},
flow/.style={-Latex,black!75,thick},
meas/.style={-Latex,black!55,dashed,thick},
block/.style={
    rectangle,rounded corners=1.5mm,draw=none,
    fill=black!7,align=center,
    minimum height=9mm,minimum width=24mm,inner sep=4pt
},
plant/.style={
    rectangle,rounded corners=1.5mm,draw=none,
    fill=purple!13,align=center,
    minimum height=9mm,minimum width=20mm,inner sep=4pt
},
emsbox/.style={
    rectangle,rounded corners=1.5mm,draw=none,
    fill=teal!15,align=center,
    minimum height=9mm,minimum width=20mm,inner sep=4pt
},
relaybox/.style={
    rectangle,rounded corners=1.5mm,draw=none,
    fill=red!20,align=center,
    minimum height=9mm,minimum width=16mm,inner sep=4pt
},
marginbox/.style={
    rectangle,rounded corners=1.5mm,draw=none,
    fill=orange!20,align=center,
    minimum height=8mm,minimum width=20mm,inner sep=4pt
},
axis/.style={->,black!65,line width=0.6pt},
vtcurve/.style={teal!70!black,line width=1.25pt},
zcurve/.style={orange!90!black,line width=1.25pt},
aux/.style={black!45,dashed,line width=0.7pt},
lab/.style={font=\scriptsize,inner sep=1pt},
tiny/.style={font=\tiny,inner sep=1pt}
]

% ------------------------------------------------------------------
% Background panels
% ------------------------------------------------------------------
\begin{scope}[on background layer]
\fill[black!2,rounded corners=2mm] (-0.15,-0.50) rectangle (4.75,2.75);
\fill[red!4,rounded corners=2mm] (5.2,-0.50) rectangle (9.7,2.75);

\node[lab,text=black!55,anchor=west] at (0.05,2.5)
    {EMS / transmission side view};
\node[lab,text=red!55!black,anchor=west] at (5.45,2.5)
    {Relay / protected side view};
\end{scope}

\node[bus,font=\large] (vt) at (1.00,1.70) {$V_t$};

\node[block] (xfmr) at (3.35,1.70)
    {transformer tap $n_i$\\fixed for $t<T_{\mathrm{tap}}$};

\node[bus,font=\large] (zi) at (6.25,1.70) {$z_i$};

\node[plant] (plant) at (8.5,1.70)
    {plant /\\collector grid};

\draw[line] (vt) -- (xfmr);
\draw[line] (xfmr) -- (zi);
\draw[line] (zi) -- (plant);

\node[emsbox] (ems) at (1.00,0.1)
    {EMS / RTCA\\checks $V_t$};

\node[relaybox] (relay) at (6.25,0.1)
    {OV relay\\$V_i^{\mathrm{trip}},\,T_i$};

\node[marginbox] (margin) at (8.5,0.1)
    {$h_i=V_i^{\mathrm{trip}}-z_i^0$};

\draw[meas] (vt) -- (ems);
\draw[flow] (zi) -- (relay);
\draw[flow,red!65!black] (relay.east) -- (margin.west);

\node[lab,text=black!65] at (5.15,1.9)
    {$z_i\simeq V_t/n_i$};

\node[lab,text=black!55,align=center] at (3.35,0.6)
{tap unchanged\\inside fast window};

\draw[aux] (xfmr.south) -- ++(0,-0.42);

\end{tikzpicture}}
\caption{Relay voltage margin. The transmission-side voltage $V_t$ and the relay voltage $z_i$ differ in general. Within a short window the tap ratio is fixed, so the relay voltage can cross the overvoltage threshold while the transmission voltage is acceptable.}
\label{fig:protection_side}
\end{figure}

\subsection{Voltage control within the relay window}\label{subsec:control_authority_mechanism}

During a fast cascade, what matters is not the total reactive capability connected to the network but the part of it that can act, in the right control mode, before the relays operate. Each control is described by its mode (voltage control, reactive power control, fixed power factor, fixed power, or limiter active), its limits, and its closed-loop response. When vendor models are unavailable, grid code envelopes or measured step responses are used, and uncertain modes and envelopes enter the parameter set of Section~\ref{sec:model}.

\vspace{0.15cm}
\noindent\textit{\textbf{Synchronous machines and excitation limiters.}} An automatic voltage regulator (AVR) controls the terminal voltage with a closed-loop excitation response of typically a few hundred milliseconds, and it is in service whether or not the unit delivers its contracted reactive power. An overexcitation limiter (OEL) restricts the field current after a thermal delay. An underexcitation limiter (UEL) restricts reactive absorption to keep the machine within its stability limit during sustained high voltage. It is the relevant limiter in an overvoltage cascade, because an active UEL reduces the unit's ability to lower the voltage. UEL and OEL transitions are discrete events, as in the standard IEEE~AC and~ST exciter models.

\vspace{0.15cm}
\noindent\textit{\textbf{Voltage control obligations (P.O.~7.4).}} Under the Spanish Operational Procedure P.O.~7.4 (2000), conventional generators above 30~MW set their reactive power themselves, from a formula of the voltage at their connection point, their installed power, and a system voltage reference published by the TSO~\cite[pp.~208, 210]{final}. The TSO computed the resulting reactive power reference only afterwards, for remuneration and monthly reports, the service counted as delivered if at least 9 of the 12 samples taken in an hour met the requirement, and the procedure did not explicitly describe the required speed of the reactive power response~\cite[pp.~210--211]{final}. On the morning of 28 April, the reactive power of several of these generators matched the reference in less than 75\% of the samples, which the Expert Panel reports without determining non-compliance~\cite[pp.~211, 333]{final}. The obligation does not change the instantaneous AVR loop but sets the operating point around which the method linearizes, and the observed deviations from the reference bound the reactive contribution of these units.

\vspace{0.15cm}
\noindent\textit{\textbf{Inverter-based voltage support.}} Inverters in voltage control or droop mode provide fast reactive support, with closed-loop responses of tens to a few hundred milliseconds~\cite{lasseter2020grid,matevosyan2019grid}. Inverters at fixed power factor do not regulate the voltage, since their reactive output follows the active power through~\eqref{eq:fixed_pf}. In the method, an inverter in voltage control mode is a control, and an inverter at fixed power factor is only a source of disturbances.

\vspace{0.15cm}
\noindent\textit{\textbf{STATCOM, SVC, and HVDC.}} STATCOMs and static var compensators (SVCs) respond typically within 100~ms. An HVDC converter contributes to voltage control in AC voltage or reactive power control mode, but not in fixed active power mode, so a change of HVDC mode, as in the chronology, changes the discrete mode of the system and its control matrices.

\subsection{Model variables}\label{subsec:mechanism_variables}

Table~\ref{tab:rootcause_map} lists how each mechanism enters the model, through the disturbances, the relay voltage margin, the controls available within the window, or the uncertainty set.

\begin{table*}[ht]
\centering
\caption{Physical mechanisms and their representation in the method. Positive $\Delta q$ raises the voltage.}
\label{tab:rootcause_map}
\small
\begin{tabularx}{\textwidth}{p{0.20\textwidth}p{0.25\textwidth}X}
\toprule
\textbf{Mechanism} & \textbf{Effect} & \textbf{Representation in the method} \\
\midrule
Fixed PF ramp & Active power change causes a proportional reactive power change. & Disturbance $d_j(t)$ with $\Delta Q_j = \sigma_j \kappa_j \Delta P_j$, only if the unit does not control voltage. \\
Generation/collector trip & Removes active power and reactive absorption. & Switching event with voltage-raising component \(\Delta q_j=Q_j^{\mathrm{abs}}\) and response from the DAE model. \\
Network meshing / line energization & Adds line charging and changes the series coupling. & Change of admittance and mode, with response from the DAE model. \\
Exchange reduction & Lower branch currents reduce the series reactive losses. & Disturbance profile, with the topology unchanged unless switching is involved. \\
Shunt reactor switching & Changes reactive absorption and the local Jacobian. & Change of admittance and mode, and a control only on horizons longer than the actuation delay. \\
LFDD / pump disconnection & Reduces load and changes reactive flows. & Load disturbance $\Delta \mathbf{d}_L(t)=\mathrm{col}(\Delta P_L,\Delta Q_L)$ with response from the DAE model. \\
Transformer tap & $n_i$ is fixed during the window. & Relay voltage $z_i$ on the protected side, with the mode updated for longer windows. \\
Overvoltage protection & Defines pickup thresholds and delays. & Relay voltage \(z_i\), margin \(h_i=V_i^{\mathrm{trip}}-z_i^0\), pickup and delay logic. \\
Synchronous machine AVR & Fast voltage control with UEL/OEL limiters. & Dynamics in $\mathbf{A}_r^m, \mathbf{B}_r^m$, with limiter activation as a mode change. \\
P.O.~7.4 obligation & The reactive power reference, and deviations from it, set the operating point. & Operating point and response envelope of obligated units, with uncertainty in \(\theta\). \\
IBR voltage control vs.\ fixed PF & Voltage control mode provides fast reactive support. & Control in voltage control mode, only a disturbance at fixed PF. \\
HVDC mode & AC voltage or reactive power control provides voltage support. & Mode-dependent control matrices, with a mode change as a discrete update. \\
Missing/uncertain data & Unknown thresholds, measurement points, or modes. & Uncertainty set $\Theta$, whose bounds on disturbances and controls enter the robust certificate. \\
\bottomrule
\end{tabularx}
\end{table*}

From the present operating point, the method computes how a candidate event changes each relay voltage over the relay time window, divides this change by the remaining margin, and compares it with the voltage reduction that the available controls can provide in the same window. Section~\ref{sec:model} develops the dynamic model used for these computations.

\section{Hybrid DAE Model and Relay Voltage Response}\label{sec:model}
% =====================================================================

The relay voltage response is computed from a hybrid differential-algebraic model. The model is hybrid because relay trips, shunt switching, tap changes, control mode changes, and limiter activations change the network and device equations. The method works mode by mode, where the mode of a relay time window fixes the discrete states that cannot change within that window, while the continuous dynamics of all active devices are retained.

\subsection{Hybrid DAE model}\label{subsec:hybrid_parent}

Let $m\in\mathcal M$ be the current discrete mode. A mode specifies which assets are connected, which shunts are in service, the present tap positions, which protection devices have already operated, and which controller modes and limiters are active (for example, an AVR with or without an active UEL, an inverter in voltage control or fixed power factor mode, or an HVDC link in reactive power control or fixed power mode). For a fixed mode, the system is described by
\begin{subequations}\label{eq:hybrid_dae}
\begin{align}
\dot{\m x} &= \m f_m(\m x,\m y,\m u,\m d;\m\theta),\label{eq:hybrid_diff}\\
\m 0 &= \m g_m(\m x,\m y,\m u,\m d;\m\theta),\label{eq:hybrid_alg}\\
\m z &= \m h_m(\m x,\m y;\m\theta).\label{eq:hybrid_output}
\end{align}
\end{subequations}
Here $\m x$ contains the dynamic states of synchronous machines, exciters, governors, power system stabilizers (PSSs), UEL/OEL limiters, inverter controls, STATCOMs, SVCs, HVDC controls, dynamic loads, and other devices, and $\m y$ contains the bus voltages (in polar or rectangular form) and algebraic device variables. The vector $\m u$ contains the setpoints and switching actions available to the operator, $\m d$ the exogenous events and disturbances, including load and pump disconnection, and $\m z$ the relay voltages. The vector $\m\theta$ collects parameters that are fixed in one computation but may be uncertain, such as tap positions, protection thresholds and delays, reactive absorption estimates, controller response parameters, P.O.~7.4 response envelopes, and model error bounds. The dependence on $\m\theta$ is omitted until Section~\ref{subsec:robust}.

Protection is represented by guard conditions. A simplified overvoltage guard is
\begin{equation}\label{eq:guard_condition}
z_i(t)\ge V_i^{\mathrm{trip}}\quad\text{for relevant delay logic}\Longrightarrow m^+=\Gamma_i(m),
\end{equation}
where $\Gamma_i(m)$ disconnects the protected element and updates the network, the controllers, and the reactive absorption. For a relay without delay, the threshold crossing triggers the mode transition immediately, and a delayed relay adds the condition that the voltage stays above pickup for the delay. UEL/OEL activation and OLTC tap changes have the same form, a guard that changes $m$ when its delay condition is met. The method evaluates each window in a fixed mode, and the mode is updated between windows when a relay, limiter, tap, or switching guard has been triggered.

\subsection{Linear model of a mode}\label{subsec:linearization}

For a mode $m$, a parameter value $\m\theta^0$, and an algebraically consistent operating point $(\m x^0,\m y^0,\m u^0,\m d^0)$, linearizing~\eqref{eq:hybrid_dae} gives
\begin{subequations}\label{eq:lin_dae}
\begin{align}
\Delta\dot{\m x} & = \m A_x^m\Delta\m x+\m A_y^m\Delta\m y+
\m B_x^m\Delta\m u+
\m D_x^m\Delta\m d,\label{eq:lin_diff}\\
\m 0 & = \m G_x^m\Delta\m x+\m G_y^m\Delta\m y+
\m B_y^m\Delta\m u+
\m D_y^m\Delta\m d,\label{eq:lin_alg}\\
\Delta\m z &= \m C_x^m\Delta\m x+\m C_y^m\Delta\m y.
\label{eq:lin_out}
\end{align}
\end{subequations}
We assume that $\m G_y^m$ is nonsingular, i.e., the DAE has index one at the operating point. Eliminating $\Delta\m y$ gives the reduced model
\begin{subequations}\label{eq:reduced_ode}
\begin{align}
\Delta\dot{\m x} &= \m A_r^m\Delta\m x+\m B_r^m\Delta\m u+\m D_r^m\Delta\m d,\label{eq:red_state}\\
\Delta\m z &= \m C_r^m\Delta\m x+\m F_u^m\Delta\m u+\m F_d^m\Delta\m d,\label{eq:red_output}
\end{align}
\end{subequations}
where
\begin{subequations}\label{eq:reduced_matrices}
\begin{align}
\m A_r^m &= \m A_x^m-\m A_y^m(\m G_y^m)^{-1}\m G_x^m,\\
\m B_r^m &= \m B_x^m-\m A_y^m(\m G_y^m)^{-1}\m B_y^m,\\
\m D_r^m &= \m D_x^m-\m A_y^m(\m G_y^m)^{-1}\m D_y^m,\\
\m C_r^m &= \m C_x^m-\m C_y^m(\m G_y^m)^{-1}\m G_x^m,\\
\m F_u^m &= -\m C_y^m(\m G_y^m)^{-1}\m B_y^m,
\qquad
\m F_d^m = -\m C_y^m(\m G_y^m)^{-1}\m D_y^m.
\end{align}
\end{subequations}
Controller dynamics in $\m x$ enter through $\m A_r^m,\m B_r^m,\m D_r^m,\m C_r^m$, so AVR, inverter, STATCOM, HVDC, load, and limiter dynamics are represented as far as they appear in the full model. Saturated and unsaturated controllers, and active and inactive limiters, are different modes, and an unknown proprietary controller is represented by a conservative measured or grid code response envelope. Discrete devices are frozen within the window, which is accurate for those whose delay exceeds it. A tap changer, for example, is fixed over a relay window of less than a second but becomes a new mode on a horizon of tens of seconds. The method is thus local in mode and horizon, while the full model remains hybrid.

\subsection{Event and control responses}\label{subsec:response}

For a step change of a parameter or setpoint, $\Delta\m d(t)=\bar{\m d}_j\mathbf 1_{t\ge0}$, starting from equilibrium, the linearized relay voltage response is
\begin{equation}\label{eq:step_response_d}
\Delta\m z_j^m(t)=
\left[\m F_d^m+\m C_r^m\int_0^t e^{\m A_r^m(t-s)}\dd s\,\m D_r^m\right]\bar{\m d}_j .
\end{equation}
For a control step $\Delta\m u(t)=\bar{\m u}_k\mathbf 1_{t\ge0}$, $\m F_d^m,\m D_r^m,\bar{\m d}_j$ are replaced by $\m F_u^m,\m B_r^m,\bar{\m u}_k$. For ramps or measured profiles, the response is the convolution
\begin{equation}\label{eq:profile_response}
\Delta\m z^m(t)=\m F_\nu^m\Delta\m\nu(t)+
\m C_r^m\int_0^t e^{\m A_r^m(t-s)}\m B_\nu^m\Delta\m\nu(s)\dd s,
\end{equation}
where $\nu$ is $u$ or $d$, with the corresponding matrices. The convolution form is needed for fixed power factor ramps and balancing actions, whose effect depends on both amplitude and timing.

A switching event $j$ (a trip, a line energization, or a shunt switching) changes the mode from $m$ to $m_j$ at $t=0$. The states cannot change instantaneously, so the algebraic variables jump to the solution $\m y_j^+$ of $\m g_{m_j}(\m x^0,\m y_j^+,\m u^0,\m d^0)=\m 0$. Because this jump can be large, the method computes it exactly, by Newton's method with the states frozen, and then uses the linear response of the post-event mode,
\begin{equation}\label{eq:switch_response}
\Delta\m z_j(t)=\Delta\m z_j(0^+)+\m C_r^{m_j}\int_0^t e^{\m A_r^{m_j}(t-s)}\dd s\;\m f_{m_j}(\m x^0,\m y_j^+,\m u^0,\m d^0),
\end{equation}
where $\Delta\m z_j(0^+)=\m h_{m_j}(\m x^0,\m y_j^+)-\m z^0$ is the exact jump of the relay voltages, the reduced matrices of mode $m_j$ are evaluated at $(\m x^0,\m y_j^+)$, and $\m f_{m_j}(\m x^0,\m y_j^+,\m u^0,\m d^0)$ is the state derivative immediately after the event.

\subsection{Pickup and trip excursions}\label{subsec:exact_window}

A relay applies threshold logic to the voltage waveform. Let
\begin{equation}\label{eq:scalar_voltage_response}
p_{ij}^m(t)=e_i^\top\Delta\m z_j^m(t)
\end{equation}
be the voltage rise at relay \(i\) caused by event \(j\) in mode \(m\), where \(e_i\) selects relay \(i\). The pickup excursion over the relay window \(T_i\) is
\begin{equation}\label{eq:window_excursion}
\mathcal M_{ij}^{m,\mathrm{pk}}(T_i)=\max_{0\le t\le T_i}[p_{ij}^m(t)]_+ .
\end{equation}
For a relay without delay, pickup is the trip condition. For a definite-time relay with delay \(T_i^{\mathrm{del}}>0\), the relevant quantity is the largest rise sustained over the delay,
\begin{equation}\label{eq:dwell_excursion}
\mathcal M_{ij}^{m,\mathrm{tr}}(T_i)=
\begin{cases}
\displaystyle
\max_{0\le \tau\le T_i-T_i^{\mathrm{del}}}\;
\min_{\tau\le t\le \tau+T_i^{\mathrm{del}}} p_{ij}^m(t),
& T_i\ge T_i^{\mathrm{del}},\\[2mm]
-\infty, & T_i<T_i^{\mathrm{del}},
\end{cases}
\end{equation}
which reduces to the pickup excursion when \(T_i^{\mathrm{del}}=0\). Dividing by the margin gives the normalized pickup and trip excursions
\begin{equation}\label{eq:K_exact}
K_{ij}^{m,\mathrm{pk}}(T_i)=\frac{\mathcal M_{ij}^{m,\mathrm{pk}}(T_i)}{h_i^m},\qquad
K_{ij}^{m,\mathrm{tr}}(T_i)=\frac{[\mathcal M_{ij}^{m,\mathrm{tr}}(T_i)]_+}{h_i^m},
\end{equation}
where \(h_i^m=V_i^{\mathrm{trip}}-z_i^{0,m}\). \(K_{ij}^{m,\mathrm{pk}}<1\) certifies that event \(j\) does not pick up relay \(i\) within the window, and for a delayed relay, \(K_{ij}^{m,\mathrm{tr}}\ge1\) is the trip condition for the linearized waveform.

At $t=0^+$, before any controller responds, the response is $\Delta\m z_j^m(0^+)=\m F_d^m\bar{\m d}_j=-\m C_y^m(\m G_y^m)^{-1}\m D_y^m\bar{\m d}_j$, or the exact jump of~\eqref{eq:switch_response} for a switching event, and it determines the pickup of fast relays. A special case shows what drives propagation. With the fixed-tap relation of Fig.~\ref{fig:protection_side}, if event $j$ removes reactive absorption $Q_j^{\mathrm{abs}}$ at transmission bus $t_j$, and $S_{ij}$ is the instantaneous sensitivity of the voltage at transmission bus $t_i$ to this loss of absorption, then
\begin{equation}\label{eq:special_K}
K_{ij}^{\mathrm{pk}}(0^+)=\frac{\left[S_{ij}Q_j^{\mathrm{abs}}/n_i\right]_+}{h_i},
\end{equation}
where \(n_i\) is the turns ratio between the transmission side and the protected side. Propagation thus depends on the electrical coupling \(S_{ij}\), the disturbance size \(Q_j^{\mathrm{abs}}\), and the exposure \(1/(n_i h_i)\) of the receiving asset, while the general expression~\eqref{eq:K_exact} adds controller dynamics, relay delays, and actuator responses.

\subsection{Bounding the window maximum}\label{subsec:window_computation}

The maximum in~\eqref{eq:window_excursion} must not be underestimated. For a step event $j$ and a relay $i$,
\begin{equation}\label{eq:scalar_channel}
\varphi_{ij}(t)=e_i^\top\Delta\m z_j^m(t),
\qquad
\dot\varphi_{ij}(t)=e_i^\top\m C_r^m e^{\m A_r^m t}\m D_r^m\bar{\m d}_j .
\end{equation}
The method samples $\varphi_{ij}$ on a grid $0=t_0<t_1<\cdots<t_N=T_i$ and bounds it between the samples. On each interval $[t_\ell,t_{\ell+1}]$ of length $\Delta_\ell$ the input is linear in time, and for a step input, $\dot{\m x}(t_\ell+s)=e^{\m A_r^m s}\dot{\m x}(t_\ell)$. With a diagonal scaling matrix $\m D$ that balances $\m A_r^m$ and an upper bound $\mu$ of the logarithmic norm $\mu_2(\m D^{-1}\m A_r^m\m D)$,
\begin{equation*}
\begin{split}
|\dot\varphi_{ij}(t_\ell+s)|\le L_{ij,\ell}&:=\|\m D\m c_i\|_2\,e^{\max(\mu,0)\Delta_\ell}\,\|\m D^{-1}\dot{\m x}(t_\ell)\|_2,\\
&\qquad 0\le s\le\Delta_\ell,
\end{split}
\end{equation*}
where $\m c_i^\top$ is the row of $\m C_r^m$ of relay $i$, and terms proportional to the input slope are added for ramps. On an interval where $|\dot\varphi_{ij}|\le L_{ij,\ell}$, the function lies below both lines of slope $\pm L_{ij,\ell}$ through the two samples, so
\begin{equation}\label{eq:lipschitz_upper}
\mathcal M_{ij}^{m,\mathrm{pk}}(T_i)
\le
\max_{\ell=0,\ldots,N-1}\left[\frac{\varphi_{ij}(t_\ell)+\varphi_{ij}(t_{\ell+1})+L_{ij,\ell}\Delta_\ell}{2}\right]_+ .
\end{equation}
The implementation also bounds $|\ddot\varphi_{ij}|$ on each interval in the same way and uses the smaller of~\eqref{eq:lipschitz_upper} and the linear interpolation bound $\max(\varphi_{ij}(t_\ell),\varphi_{ij}(t_{\ell+1}))+\max|\ddot\varphi_{ij}|\,\Delta_\ell^2/8$. Intervals whose padding exceeds a tolerance are subdivided, down to 0.25~ms. The samples are obtained by propagating the reduced model over each grid step with the matrix exponential, which is exact for piecewise-linear inputs, and for large systems its product with the state is computed with a Krylov method (Section~\ref{subsec:scaling}).

For controls, the bound goes in the other direction, because underestimating a voltage reduction is conservative. The command of control action \(k\) is parameterized as
\begin{equation}\label{eq:control_envelope}
\Delta\m u_k(t;\alpha_k)=\alpha_k \m b_k^m(t),\qquad 0\le \alpha_k\le \bar\alpha_k ,
\end{equation}
where \(\m b_k^m(t)\) is the unit response profile of action \(k\) in mode \(m\), including its delay, ramp rate, saturation, and control mode. With \(\Delta\m z_{u_k}^m\) the relay voltage response to the unit command \(\Delta\m u_k(t;1)=\m b_k^m(t)\), the normalized control coefficient at grid time $t_\ell$ is
\begin{equation}\label{eq:H_time}
H_{ik}^m(t_\ell)=
-\frac{e_i^\top\Delta\m z_{u_k}^m(t_\ell)}{h_i^m},
\qquad
\underline H_{ik}^m(t_\ell)\le H_{ik}^m(t_\ell),
\end{equation}
where \(\underline H_{ik}^m\) is a lower bound over the grid interval obtained in the same way as~\eqref{eq:lipschitz_upper}. Positive \(H_{ik}^m(t_\ell)\) means that control \(k\) lowers relay voltage $i$ at time \(t_\ell\). The mitigation problem of Section~\ref{subsec:mitigation} uses these time-resolved coefficients, so that disturbance and control effects are compared at the same instants.

\section{Cascade Screening and Mitigation}\label{sec:cascade}
% =====================================================================

The matrices of Section~\ref{sec:model} describe the effect of a single event. In a cascade, the disconnection of each tripped element is a new event that may trip others.

\subsection{Threshold cascade map}\label{subsec:threshold_map}

Let \(\mathcal C=\{1,\ldots,n_c\}\) be the set of protected generation, collector, transformer, or interface elements. For a fixed mode \(m\), let \(S\subseteq\mathcal C\) be the set of elements that have tripped. Let \(\bar r_i^{0,m}\in[0,1)\) be a normalized margin reserved for uncertainty, unmodeled voltage drift, or planned actions not yet included in the operating point. The present voltage is already included in \(h_i^m=V_i^{\mathrm{trip}}-z_i^{0,m}\), so \(\bar r_i^{0,m}\) must not count it again, and \(\bar r_i^{0,m}=0\) if all planned actions are included in the operating point and uncertainty is carried by \(\bar K\).

With an upper-bounded pickup matrix \(\bar K^{m,\mathrm{pk}}\), define
\begin{equation}\label{eq:cascade_map}
\Phi_m(S)=S\cup\left\{i\in\mathcal C\setminus S:\bar r_i^{0,m}+\sum_{j\in S}\bar K_{ij}^{m,\mathrm{pk}}\ge1\right\}.
\end{equation}
For the linearized response, this additive map is exact when the individual responses peak at the same time at the relay and conservative when they are staggered or partly cancel, so, to the extent that the individual responses superpose, it overestimates pickup propagation. For definite-time relays, a trip is confirmed from the combined waveform of the elements reached, not by adding the individual trip excursions. The fixed point result below holds for a fixed matrix \(\bar K^{m,\mathrm{pk}}\). The matrix can be recomputed when a reached trip set changes the mode, or bounded over the admissible mode changes. If only the matrix of the initial mode is used, the method ranks the events, and its largest fixed points are checked with nonlinear time-domain simulation (TDS).
\begin{myprs}[Finite convergence of threshold cascade]\label{prop:fixed_point}
For any initial set $S_0\subseteq\mathcal C$ of tripped elements, the iteration
\begin{equation}\label{eq:cascade_iter}
S_{\ell+1}=\Phi_m(S_\ell),\qquad \ell=0,1,2,\ldots,
\end{equation}
converges in at most $n_c$ iterations to the least fixed point of $\Phi_m$ that contains $S_0$.
\end{myprs}
\begin{proof}
If $S\subseteq S'$, then $\sum_{j\in S}\bar K_{ij}^{m,\mathrm{pk}}\le\sum_{j\in S'}\bar K_{ij}^{m,\mathrm{pk}}$ because $\bar K_{ij}^{m,\mathrm{pk}}\ge0$. Hence $\Phi_m(S)\subseteq\Phi_m(S')$, so $\Phi_m$ is monotone. Also $S\subseteq\Phi_m(S)$ by definition. Therefore the sequence $S_0\subseteq S_1\subseteq\cdots$ is increasing. Since $\mathcal C$ has $n_c$ elements, at most $n_c$ strict inclusions can occur. The terminal set is a fixed point, and by monotonicity it is the least fixed point containing $S_0$.
\end{proof}

\begin{mycor}[No secondary trip certificate]\label{cor:no_secondary}
If
\begin{equation}\label{eq:no_secondary}
\bar r_i^{0,m}+\sum_{j\in S_0}\bar K_{ij}^{m,\mathrm{pk}}<1,
\qquad \forall i\in\mathcal C\setminus S_0,
\end{equation}
then no protected element outside \(S_0\) picks up under the additive map~\eqref{eq:cascade_map}, and hence no secondary overvoltage trip occurs. If the same inequality holds for every set reached by~\eqref{eq:cascade_iter}, the pickup cascade stops at \(S_0\).
\end{mycor}

Proposition~\ref{prop:fixed_point} gives, for each initiating event, the set of protected elements predicted to pick up. For relays without delay, this is the predicted trip set, and for delayed relays, the elements reached are checked with the trip condition~\eqref{eq:dwell_excursion}. Initiating events can be ranked by the size of their fixed points or by the largest normalized excursion they cause at the other relays, $K_j^{\max}=\max_{i\notin\mathcal S_j}\bar K_{ij}^{m,\mathrm{pk}}$, where $\mathcal S_j$ is the set of elements that event $j$ disconnects.

\subsection{Robust screening with missing or uncertain data}\label{subsec:robust}

The quantities in~\eqref{eq:cascade_map} depend on the parameters $\m\theta$ of~\eqref{eq:hybrid_dae}, which include tap positions, reconstructed relay voltages, thresholds, delays, reactive absorption, controller modes, P.O.~7.4 response envelopes, response models, and linearization error. For an uncertainty set \(\Theta\), define
\begin{equation}\label{eq:robust_K}
\begin{split}
&\bar K_{ij}^{m,\mathrm{tr}}=\sup_{\m\theta\in\Theta}K_{ij}^{m,\mathrm{tr}}(T_i;\m\theta),\quad
\bar K_{ij}^{m,\mathrm{pk}}=\sup_{\m\theta\in\Theta}K_{ij}^{m,\mathrm{pk}}(T_i;\m\theta),\\
&\qquad\qquad\quad\quad\bar r_i^{0,m}=\sup_{\m\theta\in\Theta}r_i^{0,m}(\m\theta).
\end{split}
\end{equation}
When \(z_i\) is reconstructed, the margin \(h_i^m(\m\theta)=V_i^{\mathrm{trip}}(\m\theta)-z_i^{0,m}(\m\theta)\) also depends on \(\m\theta\). The suprema in~\eqref{eq:robust_K} are taken over parameter values with \(h_i^m(\m\theta)>0\), and if \(\Theta\) does not give a positive lower bound on \(h_i^m\), the asset is classified as having insufficient data instead of being certified.
The robust screening uses \(\bar K^{m,\mathrm{pk}}\) and \(\bar r\) in~\eqref{eq:cascade_map}. If
\begin{equation}\label{eq:robust_cert}
\bar r_i^{0,m}+\sum_{j\in S}\bar K_{ij}^{m,\mathrm{pk}}\le1-\epsilon,
\qquad \forall i\notin S,
\end{equation}
then no protected element outside \(S\) picks up, and hence no overvoltage relay trips, for any parameter in \(\Theta\), with margin \(\epsilon\). For sets that violate~\eqref{eq:robust_cert}, delayed trips are confirmed from the combined waveform with~\eqref{eq:dwell_excursion}. The bounds are computed with the procedure of Section~\ref{subsec:window_computation}, combined with interval or scenario sweeps over relay settings, tap positions, disturbance sizes, controller response times, and response envelopes. Missing relay or controller data widen \(\Theta\) and make the certificate harder to obtain. In the case studies, \(\Theta\) bounds the relay thresholds, the taps, the reconstruction error, and the size of the plant trip responses, and the certificates hold for the linearized model of each mode.

\subsection{Mitigation}\label{subsec:mitigation}

The method also computes fast corrective actions that restore the margins after the event. The mitigation constraints are imposed on the grid \(\mathcal T_i=\{t_0,\ldots,t_N\}\subset[0,T_i]\) used for the disturbance responses, with the lower-bounded control coefficients \(\underline H_{ik}^m(t_\ell)\) of~\eqref{eq:H_time}. Each candidate action has a nonnegative magnitude \(\alpha_k\in[0,\bar\alpha_k]\), so an actuator that can move in both directions is represented by two actions. For a set \(S\) of initiating events, let \(\bar m_i^S(t_\ell)\) be an upper bound satisfying
\begin{equation}\label{eq:m_seed_bound}
\bar m_i^S(t_\ell)\ge
\sup_{\m\theta\in\Theta}
\frac{
\left[
e_i^\top
\sum_{j\in S}\Delta\m z_j^m(t_\ell;\m\theta)
\right]_+
}{
h_i^m(\m\theta)
},
\qquad h_i^m(\m\theta)>0 .
\end{equation}
If the events in \(S\) occur at different times, their responses are shifted by their event times before summation, and if the timing is unknown, the sum of the individual peak bounds is conservative. With a control cost vector \(c\ge0\) and a slack penalty \(\rho>0\), the mitigation problem is the linear program (LP)
\begin{subequations}\label{eq:mitigation_lp}
\begin{align}
\min_{\m\alpha,\,\eta}\quad & c^\top\m\alpha+\rho\eta,\label{eq:mit_obj}\\
\mathrm{s.t.}\quad &
\bar r_i^{0,m}+\bar m_i^S(t_\ell)-\sum_k \underline H_{ik}^m(t_\ell)\alpha_k\le 1-\epsilon+\eta,
\label{eq:mit_margin}\\[-3.5mm]
&\hspace{36mm} \forall i\notin S,\; t_\ell\in\mathcal T_i,\nonumber\\
& 0\le\alpha_k\le\bar\alpha_k,\qquad \forall k,\label{eq:mit_limits}\\
& \eta\ge0.\label{eq:mit_slack}
\end{align}
\end{subequations}
If \(\eta^\star=0\), the selected controls keep the normalized voltage rise of every constrained relay below \(1-\epsilon\) at the grid times, for all parameters in \(\Theta\). A relay with insufficient data has no positive lower bound on its margin and is left out of~\eqref{eq:mit_margin}, so \(\eta^\star=0\) gives no guarantee for it and assumes that it does not trip. In the implementation, \(\bar m_i^S(t_\ell)\) and \(\underline H_{ik}^m(t_\ell)\) are the upper and lower bounds of~\eqref{eq:lipschitz_upper} over the grid interval that starts at \(t_\ell\), so \(\eta^\star=0\) certifies the whole window, not only the grid times. If \(\eta^\star>0\), no combination of the candidate controls can be certified to lower the relay voltages enough within the window.

\begin{myrem}[Useful versus available reactive power]\label{rem:not_reserve}
The mitigation constraint measures reactive support at the relay voltage and within the relay window. A reserve that is electrically remote, delayed, blocked by a UEL/OEL or current limit, unavailable in the active control mode, or on the wrong side of a transformer contributes little to \(\underline H_{ik}^m(t_\ell)\). The LP therefore selects controls by their effect on the relay voltages within the window, not by their MVAr rating.
\end{myrem}

\subsection{Screening algorithm}\label{subsec:algorithm}

Algorithm~\ref{alg:operational_screen} summarizes the screening. It can be run in operational planning, before switching actions, or after each state estimate.

\LinesNotNumbered
\begin{algorithm}[t]
\caption{Protection-aware dynamic voltage security screening}
\label{alg:operational_screen}
\noindent\begin{minipage}{\dimexpr\linewidth-26pt\relax}
\begin{algorithmic}[1]
\Require EMS/state estimator snapshot, protection data or bounds, candidate events, candidate controls, dynamic models or response envelopes, and uncertainty set \(\Theta\).
\Ensure Ranked initiating events and relays at risk, cascade prediction, mitigation feasibility, selected actions \(\alpha_k^\star\), and residual slack \(\eta^\star\).
\State Read the current EMS/state estimator snapshot, with the topology, voltages, flows, shunt status, generator and IBR outputs, transformer taps, HVDC mode, controller modes, limiter status, and planned operator actions.
\State Read or bound the protection data, which are the relay measurement point \(z_i\), threshold \(V_i^{\mathrm{trip}}\), delay logic \(T_i^{\mathrm{delay}}\), and the asset disconnected by each relay.
\State Select candidate events \(d_j\), such as generator and collector trips, fixed PF ramps, shunt switching, line energizations, exchange reductions, load and pump disconnection, HVDC mode changes, and N-1/N-2 contingencies.
\State Select candidate controls \(u_k\), such as AVR voltage reference changes, IBR voltage control mode, STATCOM, SVC, HVDC reactive power commands, automatic shunts, ramp smoothing, and blocking of topology actions.
\State For the current mode \(m\), linearize the hybrid DAE and compute \(\bar K^{m,\mathrm{pk}}\), \(\bar K^{m,\mathrm{tr}}\), and \(\underline H^m(t_\ell)\) with the bounds of Section~\ref{subsec:window_computation}.
\State Iterate~\eqref{eq:cascade_iter} for each initiating event or set of events, and rank the events by their fixed points and normalized excursions.
\State Solve the mitigation LP~\eqref{eq:mitigation_lp} for the high-risk events. Report feasibility, the selected actions \(\alpha_k^\star\), the required fast MVAr, and the slack \(\eta^\star\).
\State Simulate the highest-risk events and those closest to infeasibility with nonlinear RMS simulation.
\end{algorithmic}
\end{minipage}
\end{algorithm}

% =====================================================================

\section{Operational Use, Assumptions, and Scalability}\label{sec:operations}
% =====================================================================

The method is an advisory tool between the EMS with RTCA and full nonlinear dynamic simulation. After each state estimate, or before a planned switching or control action, it ranks the events that can use up relay voltage margins and checks whether fast controls can restore these margins in time. It complements RMS simulation, oscillation monitoring, and protection engineering.

\subsection{Operational use}\label{subsec:operational_use}

The method uses the network, device, and controller data of operational planning and dynamic security assessment, and the operating point of the EMS state estimate, including tap positions and controller modes. For each protected asset, it also needs the relay measurement point, the threshold, the delay, and the asset that the relay disconnects, which are kept in the protection settings database. If the relay voltage is not measured, it is computed with~\eqref{eq:pq_reconstruction} from the transmission voltage, the tap, the transformer impedance, and the plant's active and reactive power measured by the supervisory control and data acquisition (SCADA) system.

The event library is maintained like the contingency list of RTCA. It contains the trip and a fixed power factor ramp of each protected plant, load and pump disconnections, generator trips, and the switching of shunts and lines, and it is updated with planned switching, scheduled ramps, phasor measurement unit (PMU) alarms, and areas that recently operated at high voltage. Events outside the library are not screened. Quantities that the operator does not know precisely enter the uncertainty set $\Theta$ as bounds, for example the reactive power absorbed by a plant whose power factor setting is uncertain, or the output of distributed generation that the EMS does not observe, aggregated at the supplying substations. A relay whose setting or measurement point is unknown cannot be certified, and a relay that is missing from the settings database is not screened.

For each event, every protected asset that the event does not disconnect is assigned one of four classes: \emph{pickup predicted} if its score \(\bar r_i^{0,m}+\sum_{j\in S}\bar K_{ij}^{m,\mathrm{pk}}\) is at least one (\emph{trip predicted} if the trip condition~\eqref{eq:dwell_excursion} also holds), \emph{close to pickup} if the score lies in \([1-\epsilon,1)\), \emph{certified} if it is below \(1-\epsilon\), and \emph{insufficient data} if the data give no positive lower bound on its margin. A relay with insufficient data is not propagated in the cascade, so the classes of the other relays assume that it does not trip. An event is certified to cause no secondary trip when all these assets are certified. The operator receives the ranked events and relays at risk, the predicted cascades, and the fast reactive actions that keep the relay voltages below pickup, or the slack \(\eta^\star\) when no such action exists (Fig.~\ref{fig:workflow}).

The results can also be used to set up special protection schemes (SPS), also called remedial action schemes~\cite{anderson1996industry}. For each event that is not certified, a scheme is armed. It is event-based, triggered by the breaker status or setpoint change of the event as soon as it occurs, with the voltage at the relay at risk as a response-based backup. Its actions are those selected by the mitigation LP~\eqref{eq:mitigation_lp}, such as reactive power commands to STATCOMs, HVDC converters, and IBRs, a change of IBRs to voltage control, or fast shunt switching, and if they are applied within the delays assumed in the LP, $\eta^\star=0$ certifies the scheme over the relay window for every relay with sufficient data. If $\eta^\star>0$, no fast action can be certified, and the event needs a preventive action before it occurs, such as postponing a switching action, smoothing a ramp, or reconnecting a reactor. The schemes are updated after each state estimate.

\begin{figure}[t]
\centering
\resizebox{\columnwidth}{!}{%
\begin{tikzpicture}[
>=Stealth,
font=\footnotesize,
inputbox/.style={
    rectangle,rounded corners=1.5mm,draw=none,fill=black!7,
    align=center,minimum width=25mm,minimum height=9mm,inner sep=3pt
},
protbox/.style={
    rectangle,rounded corners=1.5mm,draw=none,fill=red!14,
    align=center,minimum width=25mm,minimum height=9mm,inner sep=3pt
},
modelbox/.style={
    rectangle,rounded corners=1.5mm,draw=none,fill=purple!12,
    align=center,minimum width=25mm,minimum height=9mm,inner sep=3pt
},
screenbox/.style={
    rectangle,rounded corners=1.8mm,draw=teal!60!black,line width=0.8pt,
    fill=teal!8,align=center,minimum width=25mm,minimum height=10mm,inner sep=3pt
},
outbox/.style={
    rectangle,rounded corners=1.5mm,draw=none,fill=orange!22,
    align=center,minimum width=25mm,minimum height=9mm,inner sep=3pt
},
validbox/.style={
    rectangle,rounded corners=1.5mm,draw=none,fill=purple!15,
    align=center,minimum width=25mm,minimum height=8mm,inner sep=3pt
},
sig/.style={-Latex,black!75,thick},
datasig/.style={-Latex,black!55,thick},
dashsig/.style={-Latex,black!55,dashed,thick},
lab/.style={font=\scriptsize,inner sep=1pt},
tag/.style={
    circle,fill=teal!75!black,text=white,font=\scriptsize,
    minimum size=5mm,inner sep=0pt
}
]

% ==============================================================
% Coordinates
% ==============================================================
\coordinate (xin) at (0,0);
\coordinate (xmid) at (3.4,0);
\coordinate (xout) at (6.8,0);

% ==============================================================
% Input boxes
% ==============================================================
\node[inputbox] (ems)    at (xin |- 0, 1.85) {EMS snapshot\\topology, $V,\theta,P,Q$};
\node[protbox]  (prot)   at (xin |- 0, 0.60) {protection data\\$z_i,V_i^{\mathrm{trip}},T_i$};
\node[inputbox] (events) at (xin |- 0,-0.65) {events and controls\\$d_j,u_k$, limits};
\node[modelbox] (models) at (xin |- 0,-1.90) {dynamic models\\or response envelopes};

\begin{scope}[on background layer]
\node[
    fill=black!3,rounded corners=2mm,
    fit=(ems)(prot)(events)(models),
    inner xsep=1mm, 
    inner ysep = 1.75mm,
] (inputpanel) {};
\end{scope}

\node[lab,text=black!55,anchor=south] at (inputpanel.north) {inputs};

% ==============================================================
% Screening boxes
% ==============================================================
\node[screenbox] (lin)  at (xmid |- 0, 1.45) {linearization of\\the current mode};
\node[screenbox] (maps) at (xmid |- 0, 0.00) {window bounds\\$\bar K^{\mathrm{pk/tr}}(T)$, $\underline H(t)$};
\node[screenbox] (mit)  at (xmid |- 0,-1.45) {cascade + mitigation\\$\Phi(S)$, LP};

\node[tag] at (lin.north east)  {1};
\node[tag] at (maps.north east) {2};
\node[tag] at (mit.north east)  {3};

\begin{scope}[on background layer]
\node[
    fill=teal!4,rounded corners=2mm,
    fit=(lin)(maps)(mit),
    inner xsep=1mm,inner ysep=1.7mm
] (screenpanel) {};
\end{scope}

\node[lab,text=teal!55!black,anchor=south] at (screenpanel.north) {protection-aware screening};

% Main screen flow
\draw[sig] (lin) -- (maps);
\draw[sig] (maps) -- (mit);

% Clean horizontal input arrows into each screen layer
\draw[datasig] (inputpanel.east |- lin.west)  -- (lin.west);
\draw[datasig] (inputpanel.east |- maps.west) -- (maps.west);
\draw[datasig] (inputpanel.east |- mit.west)  -- (mit.west);

% ==============================================================
% Uncertainty set
% ==============================================================
\node[
    protbox,
    minimum width=35mm,
    fill=red!11
] (unc) at (xmid |- 0, 3.05)
{uncertainty set $\Theta$\\taps, relays, modes};

\draw[dashsig] (unc.south) -- (lin.north);

% ==============================================================
% Output boxes
% ==============================================================
\node[outbox] (risk) at (xout |- 0, 0) {risk ranking\\events, relays at risk};
\node[outbox] (act)  at (xout |- 0,-1.1) {corrective action\\$\Delta u^\star$, fast MVAr};
\node[outbox] (cert) at (xout |- 0,-2.2) {certificate / slack\\no trip or $\eta^\star>0$};

\begin{scope}[on background layer]
\node[
    fill=orange!5,rounded corners=2mm,
    fit=(risk)(act)(cert),
    inner xsep=1mm,  inner ysep=1mm
] (outpanel) {};
\end{scope}

\node[lab,text=orange!80!black,anchor=south] at (outpanel.north) {outputs};

% Direct risk arrow from finite window maps
\draw[sig] (maps.east) -- (risk.west);

% One clean branch from mitigation to two outputs
\coordinate (branch) at ([xshift=5mm]mit.east);
\draw[sig] (mit.east) -- (branch);
\draw[sig] (branch) |- (act.west);
\draw[sig] (branch) |- (cert.west);

% ==============================================================
% RMS validation
% ==============================================================
\node[validbox] (rms) at (xmid |- 0,-2.95) {nonlinear RMS validation};

\draw[dashsig] (mit.south) -- (rms.north);

\node[
    align=center,
    font=\scriptsize,
    text=black!55
] at (7.05,-3.3)
{highest risk or closest-\\to infeasible cases};

\draw[dashsig] (cert.south) |- (rms.east);

\end{tikzpicture}}
\caption{Workflow of the method, between EMS/RTCA and nonlinear dynamic simulation. The matrices \(\bar K^{\mathrm{pk}}(T)\), \(\bar K^{\mathrm{tr}}(T)\) and the coefficients \(\underline H(t_\ell)\) are internal, and the operator receives ranked initiating events and relays at risk, corrective actions, the residual slack, and the assets with insufficient data.}
\label{fig:workflow}
\end{figure}

\subsection{Assumptions}\label{subsec:assumptions}

The method makes five assumptions.

\noindent\textit{A1. Regular operating point.} The linearization requires a state-estimated operating point with a nonsingular algebraic Jacobian $\m G_y^m$, as do many operational sensitivity tools. This suits screening, but not exact prediction after many trips, islanding, or loss of synchronism.

\noindent\textit{A2. Fixed mode within a window.} Shunts, taps, protection states, limiters, and control modes are fixed within one window, which is appropriate when their actuation delay or activation time exceeds the window. For longer horizons, the mode has to be updated and the matrices recomputed.

\noindent\textit{A3. Controllers are represented by models or response envelopes.} AVR, IBR, STATCOM, SVC, HVDC, UEL/OEL, and load dynamics enter through $\m A_r,\m B_r,\m D_r$, or through conservative response envelopes when detailed models are unavailable.

\noindent\textit{A4. Protection data are known or bounded.} When $V_i^{\mathrm{trip}}$, $T_i^{\mathrm{delay}}$, or the measurement point are missing, the robust screening uses intervals. Missing and heterogeneous protection and third-party data were among the difficulties of the investigation~\cite[pp.~7--8, pp.~159--164]{final}.

\noindent\textit{A5. The event library is finite.} As in RTCA, events outside the library (Section~\ref{subsec:operational_use}) are not screened.

Under these assumptions, the method screens events, and the highest-risk cases are checked with nonlinear RMS simulation.

\subsection{Scalability}\label{subsec:scaling}

The computation uses sparse linear solves and the propagation of selected relay voltages, not a nonlinear simulation of every event. For each mode, the main costs are solves with $\m G_y^m$ and products of $e^{\m A_r^m t}$ with selected vectors. Operators need only the rows and columns of \(\bar K^{\mathrm{pk}},\bar K^{\mathrm{tr}}\in\R^{p\times q}\) and \(\underline H(t_\ell)\in\R^{p\times r}\), where \(p\) is the number of relays, \(q\) the number of candidate events, and \(r\) the number of controls, that belong to high-voltage zones, electrically close plants, planned switching areas, and events with a large reactive power impact. For large systems, four reductions can be used: \textit{(i)} grouping protected assets by voltage control zone or collector area, \textit{(ii)} preselecting pairs of events and relays by electrical distance, \textit{(iii)} computing only the largest impacts with sparse adjoint solves, and \textit{(iv)} simulating only the highest-risk events.

For the 2000-bus system, the implementation avoids dense matrix exponentials and dense factorizations of the DAE Jacobian, whereas the benchmarks use dense exponentials of the reduced matrices. It removes the states that cannot affect any relay voltage, forms the reduced matrices~\eqref{eq:reduced_matrices} of each mode with a sparse LU factorization of $\m G_y^m$~\cite{davis2010klu}, and samples the responses with a shift-and-invert Krylov approximation of the matrix exponential~\cite{moret2004rd,vandeneshof2006preconditioning}, whose linear solves reuse one sparse LU factorization of the DAE Jacobian per mode. The Krylov dimension is increased until two successive approximations agree at every grid point. The logarithmic norm of Section~\ref{subsec:window_computation} is computed with Lanczos iterations and certified by a Cholesky factorization of $\mu\m I-(\hat{\m A}+\hat{\m A}^\top)/2$, $\hat{\m A}=\m D^{-1}\m A_r^m\m D$, which exists only if $\mu$ exceeds every eigenvalue. The reduced matrix of the retained states is still dense, so its cost grows with the square or the cube of their number, and much larger systems also need reductions \textit{(i)}--\textit{(iv)}.

\section{Case Studies}\label{sec:case_studies}

The method is validated against nonlinear time-domain simulation (TDS) with relays on four benchmark systems and on a 2000-bus system built from the synthetic ACTIVSg2000 case with added protected renewable plants. All simulations use ANDES~1.9.3~\cite{andes}. \ref{app:systems} gives the details of the test systems, and the code and results are publicly available~\cite{codes}.

\subsection{Benchmark systems and settings}\label{subsec:case_numerical_setup}\label{subsec:case_setup}

The benchmarks are the Kundur two-area system, the IEEE 39-bus system, the NPCC 140-bus system, and GBnetwork, a 2224-bus model of the Great Britain transmission system (Table~\ref{tab:benchmark_systems}). Each is modified in the same way to create the high-voltage condition of Section~\ref{sec:mechanism}. Loads are scaled to 60\% of their base values, and renewable energy source (RES) plants are added with a total output of 50\% of the scaled load, or of the largest share that the synchronous units allow at their minimum output, which is 13\% in Kundur and 39\% in GBnetwork. Each plant is a WECC PVD1 model at fixed power factor 0.95 absorbing. It is connected at a 34.5~kV collector bus and through a fixed-tap step-up transformer to a bus of 200~kV or more, with taps that place most plants 0.008--0.04~pu below their overvoltage pickup. Each collector has an overvoltage relay with 1.10~pu pickup and 50~ms definite-time delay that disconnects the plant, and its timer resets when the voltage falls below pickup.

The event library of each system (Table~\ref{tab:numerical_settings}) contains the trip and a fixed power factor ramp of every plant, the disconnection of the loads with the largest reactive power, trips of up to six of the largest synchronous machines with their controllers, export reductions between areas where units have headroom above their minimum output, and the switching of added reactors, capacitors, and parallel circuits of the two extra-high-voltage lines with the largest charging. The controls are STATCOMs at plant buses, reactive power commands of the plants, AVR setpoint changes of up to six of the largest machines (not in GBnetwork, whose machines are classical models), fast reactors, and manually switched reactors.

Every event of the library is used as an initiating event. The method computes its response over a 1~s window with a 5~ms grid, the predicted cascade~\eqref{eq:cascade_iter} with the trip condition of the 50~ms relays, and the class of every plant (Section~\ref{subsec:operational_use}) with $\epsilon=0.05$. The event is then simulated in TDS at 0.5~s from the same operating point, and the simulation continues for 1.5~s. An event is \emph{dangerous} if, in TDS, it trips at least one plant other than those it disconnects within 1~s, and later trips are reported but not scored. For each event and each plant that the event does not disconnect, a predicted trip is a true positive if the plant trips in TDS and a false positive otherwise. A certified plant that trips in TDS would be a \emph{missed trip}, the error the method must avoid. Plants with a predicted pickup but no confirmed trip, plants close to pickup, and plants with insufficient data are counted separately.

\subsection{Overvoltage cascade on a 2000-bus system}
\label{subsec:case_activsg2000_replica}

The 2000-bus system is the synthetic ACTIVSg2000 case~\cite{birchfield2016grid}, imported into ANDES from its RAW and DYR files and extended with protected renewable plants (Table~\ref{tab:replica_inventory}). It is not a model of the Iberian system. We use it to test whether the mechanism of Section~\ref{sec:mechanism} leads to a collapse on a large dynamic model with standard protection settings, and whether the method remains accurate at this size. Control area 7 (431 buses, 18.2~GW of load and 4.25~GW of net export in the data) is the affected area, and its 21 boundary branches form the cut set opened by out-of-step protection. Its loads are scaled to 63\%, approximately the ratio of the Spanish demand at the time of the event (25~GW~\cite[p.~11]{final}) to its annual peak, and its synchronous dispatch is reduced so that its net export is unchanged. Sixteen PVD1 plants with a total of 2906~MW are added at fixed power factor, and three of them, Granada (355~MW) and Badajoz (582 and 145~MW), have the sizes and reactive absorption of the first disconnections of the event~\cite[pp.~14, 148, 150]{final}. One common tap places the most exposed plant 0.01~pu below its pickup at the initial operating point.

Each device class has one protection setting taken from standards (\ref{app:activsg}). Plants and synchronous machines have overvoltage stages of 1.10~pu for 1~s and 1.20~pu for 0.16~s, together with undervoltage and frequency stages~\cite{nerc2020prc024,ieee1547,eu2016rfg}, and an overvoltage stage that has picked up resets only when the voltage falls below 0.98 times its pickup. Six underfrequency load shedding (UFLS) stages shed 7.5\% of the load each between 49.0 and 48.0~Hz~\cite{eu2017ncer}. Out-of-step protection opens the cut set when the angle between the centers of inertia of area 7 and the rest of the system has changed by 120\textdegree, and an island collapses when its median voltage stays below 0.5~pu or its frequency outside 47.5--52.5~Hz for 0.5~s, or when it has no synchronous machine left. All trips and the separation follow from the protection, and none is imposed. Three changes that preceded the cascade in Spain are applied~\cite{final}. Second circuits in parallel with the two 500~kV lines of the plant zone with the largest charging are energized at 1 and 3~s, the export is reduced by 1000~MW through governor setpoints, and the plant output is reduced by 208~MW at fixed power factor, both in 5--15~s.

\vspace{0.15cm}
\noindent\textit{\textbf{Base state.}} In the state built from the case data, the actions raise the voltage of the most exposed plant, the larger Badajoz plant, to 1.0999996~pu at the end of the 60~s simulation (Fig.~\ref{fig:replication_operator_conditions}a), and no plant trips within it. The voltage is then still rising slowly, and at the operating point after the actions, where the screening of Section~\ref{subsec:case_replica_screen} starts, it is 1.1001~pu, just above pickup. With smaller initial margins or fewer synchronous units online, up to four plants (949~MW) trip, at most two of them because of the loss of absorption of earlier trips. In none of 57 simulations of this state and its variants does a synchronous machine trip, load get shed, or the area separate.

\vspace{0.15cm}
\noindent\textit{\textbf{Stressed state.}} The base state differs from the Iberian system before the voltage rise in its renewable share, synchronous commitment, export, interconnection strength, reactive balance, voltage control, load, embedded generation, and protection. The stressed state changes these with the following rules. (i) Plants of 120--400~MW are added while the total stays within 45\% of the area's generation (its load plus the net export), giving 26 plants and 5.45~GW, all at fixed power factor. (ii) The smallest synchronous units of the area are decommitted until the online capacity is at most 1.5 times the synchronous dispatch, which leaves 12 units (10.6~GVA) online. (iii) The export is set to 0.87~GW, the exchange with France at 12:24~\cite[p.~47]{final}, by reducing the synchronous dispatch of the area and the load of the rest of the system by the same amount. (iv) Every machine without an exciter model in the data (327 machines, including 250 of the 334 units online in the data) receives the SEXS exciter with the default parameters that ANDES uses for the case's unsupported exciter records (gain 20, 1~s time constant). (v) The series impedance of the 21 cut-set branches is multiplied by a factor $k$, since Iberia is weakly coupled to the rest of Europe (inter-area mode of 0.2~Hz~\cite[p.~11]{final}), with $k=15$ unless stated otherwise. (vi) Just before the voltage rise, the plants at fixed power factor in the south of Spain absorbed 2.3~GVAr, the connected shunt reactors 1.3~GVAr, and the conventional units only 74~MVAr, while 2.7~GVAr of reactors were out of service~\cite[p.~216]{final}, and lines had been coupled to damp the oscillations, which raised the voltage~\cite[p.~12]{final}. To create this reactive surplus, the two circuits energized in the base state are closed from the start, and a capacitive susceptance of 1.6~GVAr is added at the area's 230 and 500~kV buses in proportion to their line charging. (vii) Under P.O.~7.4, each conventional unit set its reactive power itself from a formula of the voltage, and the TSO assessed the service only afterwards, on 12 samples per hour, without monitoring it close to real time~\cite[p.~210]{final}. The area's units keep their AVRs, but a slow loop (10~s time constant) shifts each AVR setpoint until the unit returns to its initial reactive output, so that the units do not follow the voltage rise, and an under-excitation limiter (1~s) holds each unit at the absorption limit of the data. (viii) The active power of all loads has a voltage exponent of 1.2 (80\% constant current and 20\% constant impedance), the summer residential value of the report~\cite[p.~292]{final}, and their reactive power is a constant impedance. (ix) Embedded PV is added at each load of the area with 15\% of its power, an assumed share (1.72~GW at 241 loads), and the load is raised so that the net load is unchanged. Its inverters trip at 1.10~pu for 1.5~s or 1.148~pu for 0.2~s (253 and 264~V~\cite[p.~281]{final}), and each load has five equal tranches that reach 1.10~pu when its bus voltage has risen by 0.01 to 0.06~pu, to represent the spread of voltages along low-voltage feeders. (x) The Granada plant also trips at 1.10~pu without delay, as the 220~kV protection of its transformer did~\cite[p.~213]{final}.

At the initial operating point, the plants absorb 1.80~GVAr, the 12 units supply a net 0.12~GVAr with 5 of them at their absorption limit, the added susceptance supplies 1.73~GVAr, and the area exports 0.86~GW. The median transmission-side voltage at the plants is 1.024~pu, or 410~kV on a 400~kV base, while in Spain the 400~kV voltages were below 420~kV~\cite[p.~12]{final}. Without the added susceptance, the units supply 2.27~GVAr, and this voltage is 0.982~pu. As in the base state, one common tap places the most exposed plant (Seville, 550~MW) 0.01~pu below its pickup. During the simulation, only the output reduction of the final minute is applied, with the plants reducing their output by 708~MW at fixed power factor over 48~s, the sum of the reductions of large (500~MW, at fixed power factor) and distributed (208~MW) renewable plants in Spain~\cite[p.~13]{final}. The 317~MW rise of the net load in the distribution grids is not imposed, because the report attributes it to trips of embedded generation and to voltage-dependent load~\cite[p.~13]{final}, both of which are in the model.

\begin{table}[t]
\centering
\caption{Outcome of the main run of the stressed state of the 2000-bus system ($k=15$) and of single changes. The columns give the time (s) and plant of the first trip, the plants tripped (MW) and the embedded PV tripped (MW) before the separation, and the time of separation (s) with its delay after the first trip in parentheses. After separation, all 26 plants and 9 to 11 synchronous units trip, and the solver stops during the collapse, except without plant overvoltage relays, where the island survives. Runs without separation last 100~s.}
\label{tab:iberian_state}
\scriptsize
\setlength{\tabcolsep}{2.6pt}
\begin{tabularx}{\columnwidth}{>{\raggedright\arraybackslash}Xllrl}
\toprule
Case & First trip & Plants & PV & Sep. \\
\midrule
Main run & 15.2 Seville & 16 (3209) & 358 & 19.2 (4.1) \\
No output reduction & -- & 0 & 0 & -- \\
No plant overvoltage relays & -- & 0 & 689 & 84.2 \\
Plants in volt-var control & -- & 0 & 58 & -- \\
Absorption preserved at trip & 15.2 Seville & 6 (1782) & 94 & 55.1 (40.0) \\
No added susceptance & 22.9 Badajoz 582 & 4 (1737) & 704 & 40.2 (17.3) \\
Constant-impedance load & 16.2 Badajoz 145 & 11 (2144) & 350 & 24.6 (8.4) \\
No embedded PV & 14.8 Seville & 26 (5189) & 0 & 20.7 (6.0) \\
AVR without the reactive loop & 19.7 Badajoz 145 & 17 (3938) & 348 & 21.9 (2.2) \\
Plant relays, 50~ms delay & 14.2 Seville & 26 (5253) & 1365 & 16.5 (2.3) \\
Initial margin 0.02~pu & 22.7 Badajoz 145 & 14 (2597) & 707 & 32.7 (9.9) \\
Initial margin 0.03~pu & 42.2 Badajoz 145 & 4 (1036) & 691 & 59.4 (17.2) \\
Initial margin 0.04~pu & 79.5 Granada & 1 (310) & 678 & 81.7 (2.3) \\
$k=12$ & 16.6 Seville & 18 (3977) & 1338 & 25.9 (9.3) \\
$k=10$ & 17.1 Seville & 6 (1800) & 644 & 35.1 (18.0) \\
Granada 0.005~pu from pickup & 7.7 Granada & 6 (1853) & 346 & 16.7 (9.0) \\
\bottomrule
\end{tabularx}
\end{table}

\begin{figure*}[t]
\centering
\includegraphics[width=0.9\textwidth]{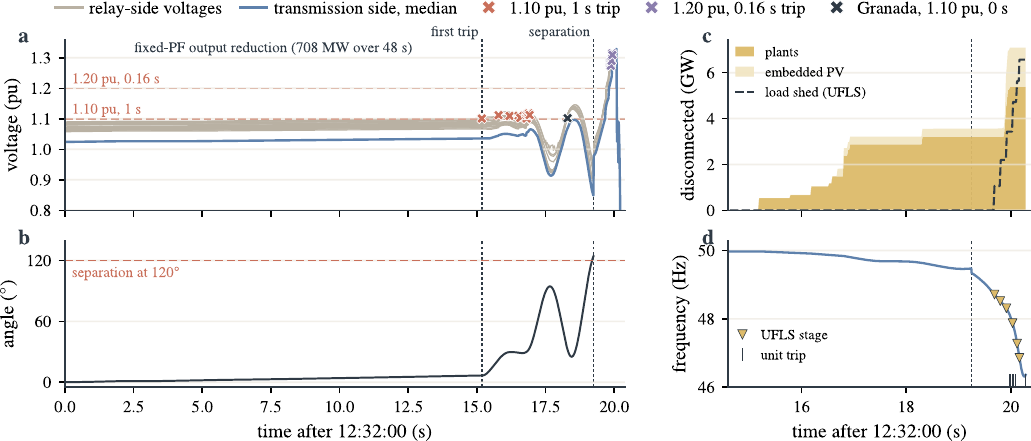}
\caption{Main run of the stressed state of the 2000-bus system ($k=15$), with time 0 at 12:32:00. (a) Relay-side voltages of the 26 plants while online, median transmission-side voltage at the plants, and plant trips by relay stage. (b) Change of the angle between the centers of inertia of the area and the rest of the system. (c) Disconnected plant and embedded PV generation, and load shed by UFLS. (d) Frequency of the system before separation and of the area after it (50~Hz equivalent), UFLS stage operations, and synchronous unit trips.}
\label{fig:replication}
\end{figure*}

Fig.~\ref{fig:replication} shows the main run. The output reduction raises the median transmission-side voltage at the plants from 1.024 to 1.034~pu in 14~s. The relay-side voltage of the Seville plant reaches 1.10~pu at 14.2~s, and the plant trips at 15.2~s. Its trip (527~MW and 173~MVAr of absorption) raises the voltages at the other plants by up to 0.011~pu, and by 16.9~s, 14 more plants have tripped on their 1~s stage, among them the Badajoz plants of 582 and 145~MW, for a total of 2.87~GW. In the same 2~s, 340~MW of embedded PV trips on overvoltage. The loss of generation swings the area against the rest, and the angle reaches 95\textdegree{} at 17.7~s, while the median transmission-side voltage dips to 0.91~pu. The Granada plant trips at 18.3~s, on the next voltage peak. At 19.2~s, 4.1~s after the first trip, the angle has changed by 124\textdegree, and out-of-step protection separates the area. The frequency of the island falls from 49.3~Hz, and the six UFLS stages shed 6.6~GW between 19.7 and 20.2~s. As load is shed, the plant-side voltages rise to 1.31~pu, the remaining 10 plants trip on their 0.16~s stage, and eight synchronous units trip on overvoltage (1.26--1.32~pu at their terminals) and one on underfrequency. The simulation stops at 20.3~s, when the solver no longer converges, with three units online, after the frequency has fallen to 46.3~Hz. The same sequence occurred on 28 April, with a voltage rise under fixed power factor operation, overvoltage trips propagated by the loss of absorption, trips of embedded PV, loss of synchronism within seconds, separation, and load shedding accompanied by overvoltage and the disconnection of the remaining generation~\cite[pp.~13--15, 21]{final}. On 28 April, the AC lines to France opened 5.1~s after the first Badajoz trip~\cite[pp.~14--15]{final}.

Table~\ref{tab:iberian_state} shows which conditions the collapse requires. Without the output reduction, nothing trips in 100~s. With the plants in volt-var control, the relay-side voltages stay below 1.099~pu, and no plant trips. The cascade needs the loss of absorption and the reactive surplus. With the absorption of tripped plants preserved, it stops after five plants (1.47~GW) within 3~s, and without the added susceptance, where the units supply 2.27~GVAr, it stops after three plants (1.42~GW), with only Granada tripping later in both cases. Without plant overvoltage relays, no plant trips, but 0.69~GW of embedded PV does. In these three cases, the area still separates, at 55.1, 40.2, and 84.2~s, but only after an inter-area oscillation has grown over many cycles, and the island survives in the last case. With the overvoltage cascade, the area separates within the first swings, a few seconds after the first trip, as on 28 April. Constant-impedance load and the absence of embedded PV slow the cascade (separation 8.4 and 6.0~s after the first trip), and 50~ms plant relays or AVR control without the reactive power loop speed it up (2.3 and 2.2~s). With larger initial margins, the area separates later, and through a growing inter-area oscillation rather than in the first swings. With 0.02 and 0.03~pu, the first plant trips at 22.7 and 42.2~s, and the area separates 9.9 and 17.2~s later. With 0.04~pu, no plant trips while the voltages rise, but an inter-area oscillation grows from about 50~s, the Granada plant trips at a voltage peak of this oscillation at 79.5~s, and the area separates at 81.7~s.

The model still differs from the Iberian event in four respects. First, the voltages rise more slowly than on 28 April, when the voltage at Granada rose by 16--17~kV in about a minute. The Expert Panel's static models also reproduce only 3--4~kV of this rise~\cite[p.~230]{final}. Without plant trips (initial margin 0.04~pu), the output reduction, the trips of embedded PV, and the voltage-dependent load raise the median transmission-side voltage at the plants by 0.025~pu (10~kV) in 48~s. With the initial margin of 0.01~pu, the first trip therefore occurs after 15~s rather than 57~s, and larger margins delay the trips but slow the cascade. Second, the Seville plant trips first, not Granada. Granada trips first when its own transformer places it 0.005~pu below pickup (the other plants then start 0.013~pu below), and on 28 April the fixed tap of its transformer did give its 220~kV side a higher per-unit voltage than the 400~kV side~\cite[p.~213]{final}. Five more plants then trip within 1.7~s, and the area separates 9.0~s after the first trip, after growing swings. Third, the interconnection is still stronger than that of Iberia, with an inter-area mode of 0.47~Hz at $k=15$ against 0.2~Hz. With $k=12$ and 10 (0.51 and 0.54~Hz), the same sequence occurs, but the cascade proceeds in bursts between growing swings, and the area separates 9.3 and 18.0~s after the first trip. Fourth, the model loses generation faster than the Iberian system did, with 2.87~GW of plants within 1.8~s of the first trip against 727 and 928~MW in the trips of 12:33:16 and 12:33:17--12:33:18~\cite[p.~14]{final}.

\subsection{Screening the 2000-bus system}\label{subsec:case_replica_screen}

\begin{figure}[t]
\centering
\includegraphics[width=0.9\columnwidth]{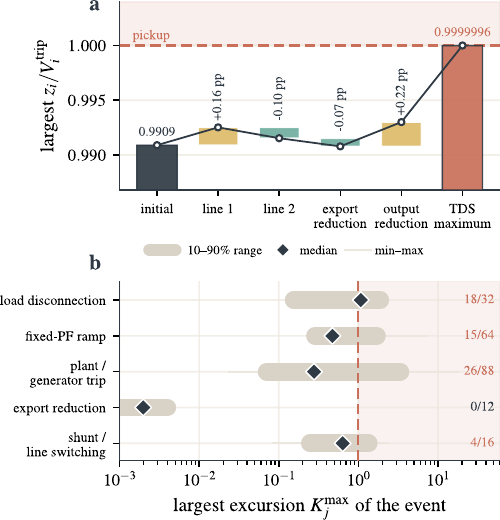}
\caption{Base state of the 2000-bus system. (a) Largest ratio of plant-side voltage to relay pickup in the power flow before and after each action, with the commitment of the case data and the most exposed plant 0.01~pu below pickup, and its maximum in TDS, with increments in percentage points. (b) Largest normalized excursion $K_j^{\max}$ of the 53 candidate events by event type in four variants of the base state (commitment of the case data or reduced commitment, and most exposed plant 0.01 or 0.02~pu below pickup before the actions) with 50~ms plant relays, and the number of events with $K_j^{\max}\ge1$ on the right. In the 0.01~pu variants, the larger Badajoz plant has insufficient data in every event and does not enter $K_j^{\max}$.}
\label{fig:replication_operator_conditions}
\end{figure}

The method was first validated in four variants of the base state after the actions, with the commitment of the case data or with an online capacity of 1.5 times the dispatch (reduced commitment), and with the most exposed plant 0.01 or 0.02~pu below pickup before the actions. The 53 events of the library are the trip and a fixed power factor ramp of each plant, 8 load disconnections, 6 generator trips, 3 export reductions, 2 capacitor switchings, and 2 line energizations. All events were simulated with a single 1.10~pu plant stage of 50~ms in the four variants and of 1~s in the two 0.02~pu variants (318 runs, with a 2~s window for the 1~s stage), and in these runs, as in the method, a relay timer resets when the voltage falls below pickup. No certified plant tripped, all 77 dangerous events had a predicted trip, and four safe events were flagged. In every dangerous event, the method identified the first secondary trip (the earliest predicted trip of the first cascade layer, which depends on the initiating event alone), at most 4~ms after and up to 0.30~s before its TDS time. The mean Jaccard index of the predicted and TDS trip sets, over the events with a predicted or TDS trip, was 0.50--0.97 per variant and relay setting, and for the 3769 plant--event pairs of events that tripped no plant, the bound was never below the TDS maximum. Over the four variants, more than half of the load disconnections reach pickup (18 of 32), and no export reduction does (Fig.~\ref{fig:replication_operator_conditions}b). In the two 0.01~pu variants, the larger Badajoz plant is above pickup at the operating point after the actions (1.1001 and 1.1081~pu), so it has insufficient data in every event and its relay is blocked in the TDS runs. With its relay active, it trips without an event after 50~ms, followed by the smaller Badajoz plant. In these variants, a generator trip has the largest excursion.

On one core, screening the 53-event library of the two variants with a 0.02~pu margin took 110 and 76~s, including the pre-event model, against 810 and 948~s for one TDS run of each event with 50~ms relays, that is, 7.4 and 12.5 times less. One screening also serves every relay setting, whereas TDS must be repeated for each. The reduced models keep 2277 to 3133 of the 4179 states, the Krylov iterations of Section~\ref{subsec:scaling} stop when two successive approximations agree to $10^{-12}$~pu, and in one variant the responses differ from those computed with dense matrix exponentials by less than $10^{-9}$~pu, while the classes and validation counts of all four variants are unchanged.

The method was then applied to the stressed state before the output reduction, with the Seville plant 0.01~pu below pickup, a state in which no plant trips in TDS without an event. Its 74 events are the trip and a fixed power factor ramp of each of the 26 plants, 8 load disconnections, 6 generator trips, 3 export reductions, 2 capacitor switchings, 2 line energizations, and the output reduction of 708~MW, applied to all plants within 0.5~s. As in the base state, the TDS runs contain only the 1.10~pu plant stage, with a 2~s window for the 1~s delay, and not the slow reactive power loop, the under-excitation limiters, or the protection of embedded PV. With the 1~s relays, which reset only below 0.98 of pickup, the state is fragile, since 33 of the 74 events are dangerous, among them the trips of 20 of the 26 plants, 7 fixed power factor ramps, and 3 of the 6 generator trips. The method flagged all 33 and 11 safe events, certified 25 events, and no certified plant tripped. Of the other five events, three are close to pickup, and the TDS runs of two, one of them for the output reduction, stop before their end and are not scored. The method identified the first trip in 30 of the 33 dangerous events, with a median timing error of 10~ms. In the other three, the plant that tripped first was predicted to pick up but not to trip, because the method assumes that a relay timer resets below pickup, whereas these relays reset only below 0.98 of pickup. Later cascade layers were overpredicted. It predicts that the output reduction drives all 26 plants above pickup ($K^{\mathrm{pk}}=9.5$) and that the Seville plant trips first, 1.23~s after the start of the reduction. In TDS with relays that reset at pickup, as in the method, Seville trips after 1.23~s and 17 plants trip within the 2~s window, and with the 0.98 reset, 23 plants trip before the solver stops. With 50~ms relays, 16 TDS runs stop during the cascade, all of them for flagged events, and in the other 58, all 19 dangerous events were flagged and no certified plant tripped. For the events that tripped no plant in TDS, the bound was never below the TDS maximum (982 and 1313 plant--event pairs for the two relay settings). On one core, screening the library took 220~s, including the pre-event model, against 4713 and 11148~s for one TDS run of each event with the 50~ms and 1~s relays, summed over runs on 8 parallel workers. Before the voltage rise, the method thus flags the output reduction and 45 of the other 73 events as dangerous, among them all 33 that are dangerous in TDS.

\subsection{Validation against nonlinear simulation}\label{subsec:case_safety}

\begin{figure*}[t]
\centering
\includegraphics[width=0.9\textwidth]{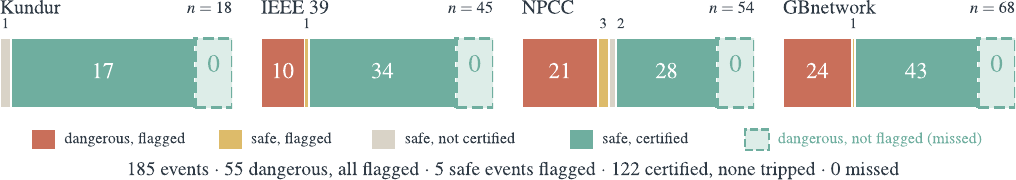}
\caption{Outcome of the 185 benchmark events in the method and in TDS (1~s window, 50~ms relays). An event is dangerous if it trips in TDS at least one plant other than those it disconnects, flagged if its predicted trip set is not empty, and certified if every other plant is certified. Missed events are dangerous events that are not flagged.}
\label{fig:validation}
\end{figure*}

Fig.~\ref{fig:validation} compares the method with TDS for all 185 events of the four benchmark libraries, and Table~\ref{tab:validation_summary} gives the totals per system. Of the 185 events, 55 are dangerous, none of them in Kundur, whose renewable share is limited to 13\%. Among the 2782 scored combinations of event and plant there is no missed trip, so no certified plant trips in TDS, and all 122 events certified to cause no secondary trip are safe. Every dangerous event has a nonempty predicted trip set, and five safe events are flagged. In six cases, a plant whose pickup was predicted without a confirmed trip did trip in TDS, but these plants were flagged, not certified. In every dangerous event, the first predicted secondary trip is the first TDS trip, with a time error of at most 15, 45, and 42~ms in IEEE 39, NPCC, and GBnetwork. In the first cascade layer, the method predicts 22, 47, and 55 trips correctly and 1, 4, and 1 incorrectly, whereas later layers are overpredicted, because pickups propagate in~\eqref{eq:cascade_map} and the trip responses are added without time shifts. Of the 264 false positives, 258 are in later layers, and the mean Jaccard index of the predicted and TDS trip sets, over the events with a predicted or TDS trip, is 0.52, 0.39, and 0.74 (0.56 overall). The false alarm with the largest excursion is a generator trip in NPCC ($K^{\mathrm{pk}}=1.25$ without a pickup in TDS), the event type that the linear model describes least accurately.

These results use the nominal screening with collector telemetry. With the telemetry error of $\pm0.002$~pu in the robust screening, there is still no missed trip, 110 events remain certified, and the false positives increase from 264 to 362. With a 2~s window, 57 events are dangerous, and there is still no missed trip, although the disconnection of load 0 in Kundur, which is certified for this window, trips all four plants 2.4 to 2.9~s after the event. Propagating only confirmed trips instead of pickups would remove 31 false positives at 1~s but would miss two trips with the 2~s window, one in Kundur and one in IEEE 39. Screening a whole library took 0.4 to 71~s, 8.8 to 42 times less than simulating it.

\begin{table*}[t]
\centering
\caption{Validation totals per system (1~s window). Counts are over combinations of an event and a plant that it does not disconnect, except Events and Dang.\ (events that trip in TDS within the window at least one plant other than those they disconnect). The columns give the true positives (TP), false positives (FP), and true negatives (TN), the plants with a predicted pickup but no confirmed trip that tripped or did not trip in TDS (Pickup), the plants close to pickup that tripped or did not trip (Close), the certified plants that tripped (Missed), the TDS trips after the window, which are not scored (Late), the mean Jaccard index of the predicted and TDS trip sets (J), and the runtimes for the whole event library (for TDS, summed over runs on six parallel workers). No plant had insufficient data, and every TDS run converged.}
\label{tab:validation_summary}
\scriptsize
\setlength{\tabcolsep}{3.4pt}
\begin{tabular*}{\textwidth}{@{\extracolsep{\fill}}l*{12}{r}}
\toprule
System & Events & Dang. & TP & FP & Pickup & Close & Missed & TN & Late & J & Method (s) & TDS (s) \\
\midrule
Kundur & 18 & 0 & 0 & 0 & 0/0 & 0/1 & 0 & 67 & 0 & -- & 0.4 & 18.7 \\
IEEE 39 & 45 & 10 & 62 & 41 & 5/2 & 0/0 & 0 & 418 & 0 & 0.52 & 7.1 & 69.1 \\
NPCC & 54 & 21 & 86 & 155 & 0/19 & 0/2 & 0 & 586 & 0 & 0.39 & 10.4 & 134.2 \\
GBnetwork & 68 & 24 & 224 & 68 & 1/20 & 0/2 & 0 & 1023 & 2 & 0.74 & 71.0 & 623.6 \\
\midrule
All & 185 & 55 & 372 & 264 & 6/41 & 0/5 & 0 & 2094 & 2 & 0.56 & & \\
\bottomrule
\end{tabular*}
\end{table*}

\subsection{Predicted cascades and event families}\label{subsec:case_cascade}\label{subsec:case_action_family}

\begin{figure*}[t]
\centering
\includegraphics[width=0.9\textwidth]{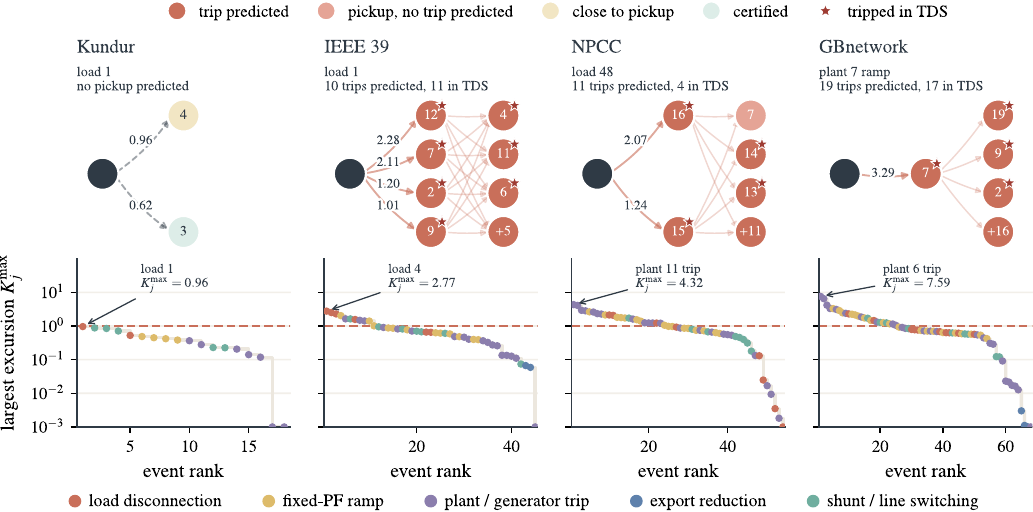}
\caption{Predicted cascades in the four benchmarks. The top row shows the event with the largest predicted cascade in each system (dark node), the plants predicted to pick up in the first layer of~\eqref{eq:cascade_iter} with $K_{ij}^{\mathrm{pk}}$ on the links, and those of later layers, where +$n$ counts the plants not shown. In Kundur, no plant is predicted to pick up, and the two plants with the largest excursions are shown. Stars mark trips in TDS within the 1~s window. The bottom row shows the largest normalized excursion $K_j^{\max}$ of every event of the library, sorted, and the dashed line marks pickup.}
\label{fig:cascade_lift}
\end{figure*}

For the event with the largest predicted cascade in each system, the method predicts the following (Fig.~\ref{fig:cascade_lift}, top). In Kundur, the largest normalized excursion of any event is 0.96, so no secondary pickup is predicted, and TDS shows no trip within the window. In IEEE 39, a load disconnection is predicted to pick up 12 plants and trip 10 of them in four layers, and TDS trips 11, including the 10 predicted. In NPCC, a load disconnection is predicted to trip 11 plants in six layers, and TDS trips 4, all of them predicted. In GBnetwork, the fixed power factor ramp of plant 7 is predicted to trip 19 plants in three layers, and TDS trips 17, 16 of them predicted. The first trip, which depends only on the response to the initiating event, agrees with TDS within 45~ms, whereas the times of later trips are estimates, since the trip responses are added without time shifts, and they are mostly earlier than in TDS. Because no certified plant tripped, every plant that tripped in TDS within the window had been predicted to pick up or classified as close to pickup, so the predicted cascade is a conservative estimate of its extent, to be confirmed by simulation.

\begin{figure}[t]
\centering
\includegraphics[width=0.9\columnwidth]{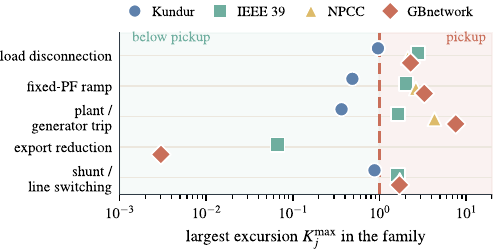}
\caption{Largest normalized excursion $K_j^{\max}$ per event type and system in the four benchmarks. The dashed line marks pickup, and events to its right are predicted to pick up at least one other plant.}
\label{fig:action_screen}
\end{figure}

Over the whole event libraries, 0 of 18, 11 of 45, 24 of 54, and 25 of 68 events reach pickup ($K_j^{\max}\ge1$, Fig.~\ref{fig:cascade_lift}, bottom), with maxima of 0.96, 2.77, 4.32, and 7.59, and all dangerous events are among them. They are plant trips, fixed power factor ramps, load disconnections, and reactor disconnections, which remove reactive absorption, together with one generator trip in NPCC and one capacitor switching in GBnetwork, while events that lower the voltage have $K_j^{\max}=0$. In IEEE 39, NPCC, and GBnetwork, each of the four families that remove absorption contains dangerous events, with largest excursions between 1.3 and 7.6 (Fig.~\ref{fig:action_screen}). Plant trips are the most severe family in NPCC, where 11 of 16 are dangerous, and in GBnetwork, where they cause the largest excursion. Generator trips, export reductions, and line energizations cause no secondary trip in TDS, and export reductions use at most 7\% of any margin. In every family, the share of events that reach pickup is equal to or slightly larger than the share of dangerous events, so the ranking of event families can be used to decide which planned actions need a detailed study.

\subsection{Mitigation within the relay window}\label{subsec:case_mitigation_window}

\begin{figure}[t]
\centering
\includegraphics[width=0.9\columnwidth]{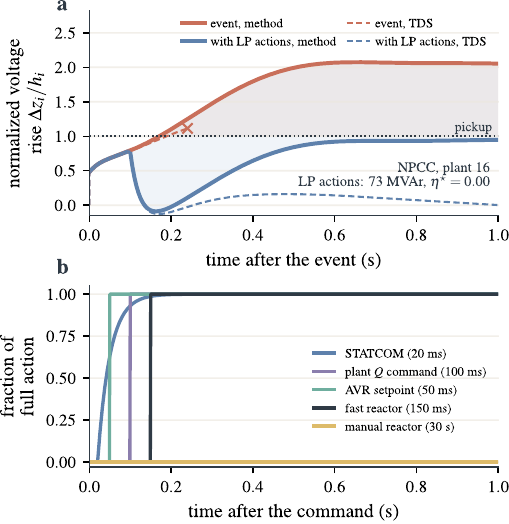}
\caption{Mitigation at the NPCC event with the largest predicted cascade (disconnection of load 48). (a) Normalized voltage rise of the most affected plant without and with the LP actions, in the method (solid) and in TDS (dashed), with crosses marking relay trips in TDS. (b) Response envelopes of the controls, with their delays.}
\label{fig:relay_window}
\end{figure}

The mitigation LP~\eqref{eq:mitigation_lp}, with \(c_k=1\) for the full range of each control and \(\rho=10^6\), was solved with all controls for every event that reaches pickup (11, 24, and 25 events in IEEE 39, NPCC, and GBnetwork), and the selected actions were then applied in TDS through their response envelopes. At the NPCC event with the largest predicted cascade (Fig.~\ref{fig:relay_window}), 73~MVAr of additional absorption by the plants' reactive power commands removes all 4 trips. At the corresponding IEEE 39 event, the load disconnection raises the voltage of plant 12 by 86\% of its margin at the instant of the event and above pickup within 5~ms, before any control can act (the fastest, the STATCOM, starts after 20~ms). The LP is therefore infeasible ($\eta^\star=0.69$), and TDS confirms the 11 trips. At the GBnetwork event, 67~MVAr removes all 17 trips within the 1~s window, but the voltages keep rising, and 12 plants trip after the window. Over all events, the LP is feasible for 4, 16, and 14 of them, and in every feasible case TDS with the LP actions shows no trip within the window. Six of the 14 GBnetwork cases, including this event, cascade after the window. Recomputed with a 2~s window, these six need 1.3 to 2.0 times the reactive power, and all 14 remain free of trips until the end of the simulation. Most infeasible cases are switching events whose instantaneous voltage jump already exceeds pickup, before any control can act.

Restricting the LP to one control class shows which reserves are useful (Remark~\ref{rem:not_reserve}). At the NPCC and GBnetwork events, only the plants' reactive power commands are sufficient on their own, and the LP uses 73 of 1434~MVAr and 67 of 2426~MVAr of this class. STATCOMs (400 and 500~MVAr), fast reactors (200 and 250~MVAr), and AVR setpoint changes (NPCC) alone are infeasible, and the manually switched reactors (400 and 500~MVAr) contribute nothing within the window. This agrees with the final report, according to which manual voltage control was too slow in the final 38~s~\cite[p.~328]{final}.

\subsection{Observability of the relay voltage}\label{subsec:case_observability}

\begin{figure}[t]
\centering
\includegraphics[width=0.9\columnwidth]{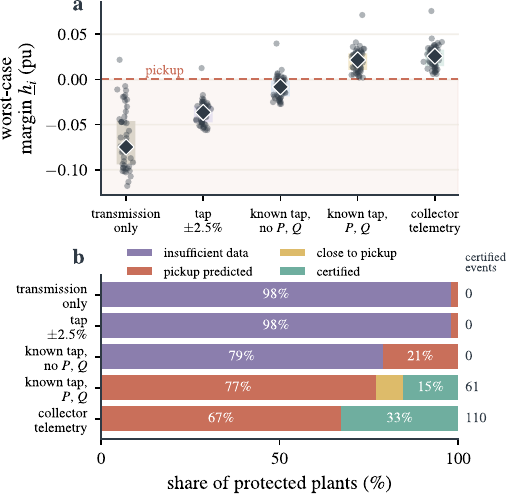}
\caption{Five levels of information on the relay voltage (the $P$, $Q$ case uses~\eqref{eq:pq_reconstruction}). (a) Worst-case margin $\underline h_i$ of every protected plant of the four benchmarks, with medians as diamonds and interquartile ranges as boxes. (b) Share of the plants per class, and events certified to cause no secondary trip (of 185).}
\label{fig:observability}
\end{figure}

Fig.~\ref{fig:observability} repeats the screening for five levels of information on the relay voltage: \textit{(i)} collector telemetry (error $\pm0.002$~pu), \textit{(ii)} a known tap, with the plant's $P$ and $Q$ measured at the transmission side and $z_i$ computed with~\eqref{eq:pq_reconstruction}, a voltage error of $\pm0.002$~pu, a power error of $\pm2\%$, and a transformer impedance error of $\pm10\%$, not less than the tolerance on the declared short-circuit impedance at the principal tap~\cite{iec60076} (the same relative error is applied to the change of the transformer drop during the event and added to the excursions, as a first-order allowance taken from the sampled response that is not part of the bound of Section~\ref{subsec:window_computation}), \textit{(iii)} a known tap without $P$ and $Q$, \textit{(iv)} a tap known within $\pm2.5\%$, and \textit{(v)} transmission-side measurements only (tap in $[0.90,1.10]$, error $\pm0.02$~pu).

In the last three cases, no voltage drop across the plant transformer can be credited, because for a plant absorbing reactive power at fixed power factor, $z_i\le V_{t_i}/n_i$ for any plant output, and this bound exceeds the actual collector voltage by the transformer drop, about 0.034~pu here, which is larger than most margins.

With collector telemetry and its error of $\pm0.002$~pu, no plant has insufficient data, and 110 of the 185 events are certified to cause no secondary trip (122 without the error, Section~\ref{subsec:case_safety}). With the known tap and measured $P$ and $Q$, $z_i$ is known within $\pm0.0066$~pu, no plant has insufficient data, and 61 events are certified (77 with a $\pm5\%$ impedance error, 24 with $\pm20\%$). Without $P$ and $Q$, 79\% of the plants have insufficient data and no event can be certified, and with a $\pm2.5\%$ tap band or transmission data only, 98\% have insufficient data. Measuring $P$ and $Q$ does not help much when the tap is uncertain (73\% with insufficient data with the $\pm2.5\%$ band). The wider bounds only make the screening more cautious than with telemetry, so every event certified in these cases is also certified with telemetry, which had no missed trip. The relay voltage must therefore be known to better than its margin, either by measuring it or by computing it from the tap, the transformer impedance, and the plant's measured power. This assumes one equivalent transformer between the plant and the transmission bus, and a plant with an internal collector network needs a wider error budget.

\subsection{Sensitivity to the uncertainty bounds}\label{subsec:case_uncertainty}

\begin{figure}[t]
\centering
\includegraphics[width=0.9\columnwidth]{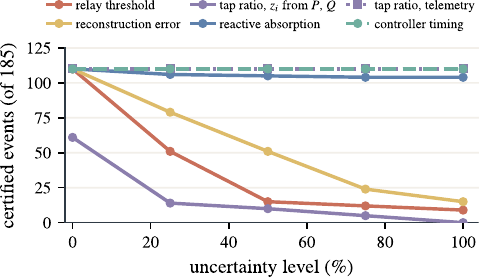}
\caption{Events certified to cause no secondary trip in the four benchmarks when one uncertainty class is widened at a time, on top of the $\pm0.002$~pu telemetry error. The tap curve with $z_i$ from $P$, $Q$ applies the tap uncertainty to $z_i$ computed with~\eqref{eq:pq_reconstruction} and starts from the 61 events certified in that case.}
\label{fig:uncertainty_value}
\end{figure}

Fig.~\ref{fig:uncertainty_value} widens one uncertainty class at a time, on top of the telemetry error, from level $\ell=0$ to $\ell=1$. The relay threshold is widened by $\pm0.02\ell$~pu, the reconstruction error by $\pm0.01\ell$~pu, the tap ratio by $\pm2.5\ell\%$ (with telemetry, and with $z_i$ computed from the measured $P$ and $Q$), the voltage response to each plant trip, as a proxy for its reactive absorption, by $\pm30\ell\%$, and the controller delays and time constants by the factor $1+\ell$. The relay threshold and the reconstruction error dominate. A threshold uncertainty of $\pm0.01$~pu ($\ell=0.5$) removes every certificate in IEEE 39 and GBnetwork and all but one in NPCC, and it lowers the Kendall $\tau$ between the ranking of the events and the nominal ranking to 0.46--0.62 in these systems. A $\pm30\%$ uncertainty of the absorption of tripping plants changes few certificates (from 29 to 28 in IEEE 39, 25 to 24 in NPCC, and 39 to 35 in GBnetwork) and keeps $\tau\ge0.90$. The tap ratio has no effect with telemetry, since the relay measures the collector voltage directly. When $z_i$ is computed from the measured $P$ and $Q$ with~\eqref{eq:pq_reconstruction}, a tap error of only $\pm0.6\%$ ($\ell=0.25$) removes every certificate in IEEE 39, NPCC, and GBnetwork, so the tap position must be known. Controller timing does not change the screening, only the mitigation LP, whose slack never decreases when the delays grow. For the other classes, the reactive power selected by the LP is not monotone in the level, because a relay whose margin bound reaches zero has insufficient data and leaves the LP. For the ramp of plant 12 in IEEE 39, it is 41.8, 61.2, and 0.1~MVAr at $\ell=0$, 0.25, and 0.5 of the threshold uncertainty.

\subsection{Protection and control changes, and ranking of events for simulation}\label{subsec:case_intervention}

\begin{figure}[t]
\centering
\includegraphics[width=0.9\columnwidth]{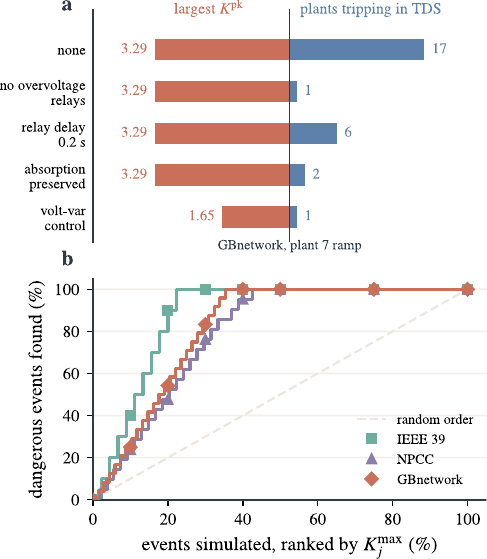}
\caption{(a) Largest $K^{\mathrm{pk}}$ in the method and plants that trip in TDS for the ramp of plant 7 in GBnetwork, with one protection or control change at a time (without overvoltage relays, the plants that reach the relay condition). (b) Share of the dangerous events found when each event library is simulated in decreasing order of $K_j^{\max}$, with random order dashed. Kundur has no dangerous event.}
\label{fig:intervention_triage}
\end{figure}

Table~\ref{tab:ablation_summary} repeats the screening and TDS for the event with the largest predicted cascade of each system, with one change at a time. In Kundur, no change produces a trip. Without overvoltage protection there is no cascade, as only 3, 1, and 1 plants in IEEE 39, NPCC, and GBnetwork reach the relay condition, compared with 11, 4, and 17 trips with protection. The cascades thus propagate through the loss of absorption at relay trips. Preserving the absorption of tripped plants, with a reactor at the collector that replaces the plant's reactive power, reduces the TDS cascades to 3, 1, and 2 trips. Operating the plants in voltage control mode (volt-var droop) reduces them to 2, 0, and 1 trips, and the method represents the droop, with the largest excursion falling from 2.28, 2.07, and 3.29 to 1.40, 1.04, and 1.65. The method follows these reductions in IEEE 39 and NPCC, but still predicts 18 trips in GBnetwork, where its later layers propagate pickups. A relay delay of 0.2~s reduces the GBnetwork cascade from 17 to 6 trips (Fig.~\ref{fig:intervention_triage}a) and has little effect in the other systems. Removing the fast reactors from the LP changes nothing, and without collector telemetry every other plant except one in NPCC has insufficient data.

Finally, Fig.~\ref{fig:intervention_triage}b ranks each event library by $K_j^{\max}$ and simulates the events in this order. The area under the receiver operating characteristic curve of this ranking is 1.000, 0.994, and 1.000 in IEEE 39, NPCC, and GBnetwork (0.999 overall), and simulating the top 22\%, 43\%, and 35\% of the libraries (31\% overall) finds every dangerous event.

\section{Black Swan or Gray Rhino?}\label{sec:answer}

The answer depends on whether one considers the exact trajectory of the event or its mechanism. The trajectory had features of a black swan, since two oscillation episodes, operator countermeasures, fixed power factor ramps near a schedule boundary, local protection behavior, missing data, and fast generation disconnections interacted in a way that no single indicator revealed, and the minute-by-minute path would have been hard to predict.

The mechanism was a gray rhino. The root cause tree of the final report lists weaknesses that were visible before the collapse, including small margins between the operating and disconnection voltages, fixed power factor operation, several conventional units whose reactive power matched the P.O.~7.4 reference in less than 75\% of the samples, no real-time monitoring by the TSO of the deviation from this reference, manual shunt reactor operation, and voltage control of local generation networks and overvoltage protection settings not aligned with system needs~\cite[pp.~331--334]{final}. These are observable weaknesses of voltage control, protection coordination, and margin management.

The proposed method checks for this mechanism during operation. A system is exposed to it when relay margins are small, plausible events can use them up through loss of reactive absorption or a topology-dependent voltage rise, and the available controls cannot lower the relay voltages within the relay window. Transmission voltages within static limits are necessary but not sufficient, and the relay margins, the effect of each event, the control response within the window, and the data uncertainty must be assessed together. In the 2000-bus system, the same weaknesses (plants at fixed power factor absorbing a reactive surplus close to their overvoltage pickup, slow reactive control of the synchronous units, and embedded PV that trips on overvoltage) turned a renewable output reduction into the collapse of the area, while volt-var control of the plants prevented any plant trip. Before the reduction, the method flagged it as dangerous, together with 45 of the other 73 events in its library.

The final report's recommendations act on the same quantities~\cite[pp.~455--460]{final}. Voltage control mode for IBRs, faster or automatic reactive power devices, visibility of dynamic reactive margins, harmonized voltage ranges, smoother active and reactive power ramps under fixed power factor, monitoring of the performance of reactive power resources, and better coordination of protection settings increase the relay margins, reduce the voltage rise caused by events, speed up the control response, or reduce the uncertainty of the certificate. The Iberian blackout thus exposed a class of operating conditions that can be screened, ranked, and mitigated before the next plausible event turns them into a cascade.

\section{Conclusion and Limitations}\label{sec:conclusion}
This paper developed a protection-aware dynamic voltage security assessment for overvoltage cascades, in which each relay trip removes reactive absorption and raises the voltage at other relays. For each candidate event, the method computes the voltage at every overvoltage relay over the relay time window, predicts which relays trip after the first trip, certifies the relays that cannot pick up, identifies those whose data are insufficient for a conclusion, and finds fast reactive power actions that keep all relay voltages below their settings.

In nonlinear simulations of four benchmark systems and a 2000-bus system, no plant that the method declared safe tripped within the relay window, every event that caused a secondary trip within it was flagged, and the plant that tripped first in TDS was always among those predicted to pick up, though not always the one predicted to trip first, while later trips were overpredicted. Simulating the events in the order ranked by the method found every dangerous event within the first half of each benchmark library, and on the benchmarks every mitigation found feasible prevented the trips within the relay window. An event can be certified to cause no secondary trip only if the relay voltages are measured, or computed from the plants' measured power with a known tap. In a stressed state of the 2000-bus system similar to Iberia before the blackout, a renewable output reduction caused a cascade with the same sequence of events as on 28 April, and the method flagged this reduction before it occurred.

The method has limitations. It screens only the events in its library, as RTCA does. It holds taps, limiters, and controller modes fixed within a window, so a limiter that activates inside the window is not represented. Its trip check assumes that a relay resets when the voltage falls below pickup. It needs protection and controller data, or tight bounds on them. Its cost grows faster than that of simulation with the size of the reduced model, so very large systems need further reductions. Finally, it was validated against simulations of the same models, which contain no OEL, UEL, OLTC, or HVDC models, and a comparison with field measurements remains to be done.

\section*{Declaration of generative AI and AI-assisted technologies in the manuscript preparation process}
During the preparation of this work, the authors used Claude and ChatGPT to assist with the implementation of the code and to improve the writing of parts of the manuscript. The authors reviewed, tested, and edited the code and the text as needed and take full responsibility for the content of the published article.

\FloatBarrier
\appendix
\setcounter{table}{0}
\setcounter{figure}{0}

\section{Test systems and additional results}\label{app:systems}

\begin{table*}[t]
\centering
\caption{Benchmark systems as simulated, with buses $n_b$ and branches $n_\ell$ (including the plant transformers and the out-of-service candidate circuits), differential states $n_x$ and algebraic variables $n_y$ of the linear model, protected plants $p$ (PVD1 models, whose relay voltages are computed rather than measured in four of the five cases of Section~\ref{subsec:case_observability}), controls $r$, candidate events $q$, and renewable share of the load.}
\label{tab:benchmark_systems}
\scriptsize
\setlength{\tabcolsep}{3.2pt}
\begin{tabular*}{\textwidth}{@{\extracolsep{\fill}}l*{8}{r}llllr}
\toprule
System & $n_b$ & $n_\ell$ & $n_x$ & $n_y$ & $p$ & $r$ & $q$ & RES (\%) & Machines & Exciters & Governors & PSS & Loads \\
\midrule
Kundur & 14 & 21 & 68 & 228 & 4 & 11 & 18 & 13 & 4 GENROU & 4 EXDC2 & 4 TGOV1 & -- & 2 \\
IEEE 39 & 51 & 60 & 238 & 771 & 12 & 27 & 45 & 50 & 10 GENROU & 10 IEEEX1 & 10 TGOV1N & 10 IEEEST & 19 \\
NPCC & 156 & 251 & 398 & 1725 & 16 & 34 & 54 & 50 & 27 GENROU, 21 GENCLS & 24 IEEEX1 & 29 TGOV1 & -- & 92 \\
GBnetwork & 2244 & 3229 & 868 & 9202 & 20 & 35 & 68 & 39 & 394 GENCLS & -- & -- & -- & 483 \\
\bottomrule
\end{tabular*}
\end{table*}

% Table A.4 is defined here so that it can share a page with Table A.1; the counter keeps the numbering in citation order.
\setcounter{table}{3}
\begin{table*}[t]
\centering
\caption{Protection and control changes at the event with the largest predicted cascade of each system (Kundur, where no change produces a trip, is omitted). The columns give the largest normalized pickup and trip excursions at the other plants ($K^{\rm pk}$, $K^{\rm tr}$), the predicted trips (Pred.), the result of the screening for the event (Class), the relay trips of other plants within 1~s (TDS), the plants whose collector voltage met the relay condition in TDS, equal to TDS when the relays are enabled (Relay cond.), the mitigation LP with all controls (LP) with its reactive power and slack, and the trips with the LP actions applied (TDS w/ LP). Removing the fast reactors from the LP gives the same results as the row ``none'' in every system. The rows without collector telemetry use transmission-side data only ($n\in[0.90,1.10]$, error $\pm0.02$~pu), and ``no certifiable plant'' means that every other plant has insufficient data.}
\label{tab:ablation_summary}
\scriptsize
\setlength{\tabcolsep}{2.6pt}
\begin{tabularx}{\textwidth}{>{\raggedright\arraybackslash}p{0.16\textwidth}>{\raggedright\arraybackslash}X*{3}{r}lrrlrrr}
\toprule
System (event) & Change & $K^{\rm pk}$ & $K^{\rm tr}$ & Pred. & Class & TDS & Relay cond. & LP & MVAr & $\eta$ & TDS w/ LP \\
\midrule
IEEE 39 (disconnection of load 1) & none & 2.28 & 2.26 & 10 & trip predicted & 11 & 11 & infeasible & 34.0 & 0.691 & 11 \\
 & no overvoltage relays & 2.28 & 2.26 & 0 & pickup predicted & 0 & 3 & infeasible & 34.0 & 0.691 & -- \\
 & relay delay 0.2~s & 2.28 & 2.17 & 10 & trip predicted & 11 & 11 & infeasible & 34.0 & 0.691 & 11 \\
 & absorption preserved & 2.28 & 2.26 & 4 & trip predicted & 3 & 3 & infeasible & 34.0 & 0.691 & 2 \\
 & plants in volt-var control & 1.40 & 1.09 & 5 & trip predicted & 2 & 2 & infeasible & 0.0 & 0.446 & -- \\
 & no collector telemetry & -- & -- & 0 & insufficient data & 11 & 11 & no certifiable plant & 0.0 & 0.000 & -- \\
\midrule
NPCC (disconnection of load 48) & none & 2.07 & 2.07 & 11 & trip predicted & 4 & 4 & feasible & 73.0 & 0.000 & 0 \\
 & no overvoltage relays & 2.07 & 2.07 & 0 & pickup predicted & 0 & 1 & feasible & 73.0 & 0.000 & -- \\
 & relay delay 0.2~s & 2.07 & 2.06 & 9 & trip predicted & 3 & 3 & feasible & 73.0 & 0.000 & 0 \\
 & absorption preserved & 2.07 & 2.07 & 3 & trip predicted & 1 & 1 & feasible & 73.0 & 0.000 & 0 \\
 & plants in volt-var control & 1.04 & 1.04 & 2 & trip predicted & 0 & 0 & feasible & 9.0 & 0.000 & 0 \\
 & no collector telemetry & 2.16 & 2.14 & 1 & trip predicted & 4 & 4 & feasible & 163.7 & 0.000 & -- \\
\midrule
GBnetwork (fixed PF ramp of plant 7) & none & 3.29 & 3.22 & 19 & trip predicted & 17 & 17 & feasible & 66.9 & 0.000 & 0 \\
 & no overvoltage relays & 3.29 & 3.22 & 0 & pickup predicted & 0 & 1 & feasible & 66.9 & 0.000 & -- \\
 & relay delay 0.2~s & 3.29 & 2.98 & 18 & trip predicted & 6 & 6 & feasible & 66.9 & 0.000 & 0 \\
 & absorption preserved & 3.29 & 3.22 & 18 & trip predicted & 2 & 2 & feasible & 66.9 & 0.000 & 0 \\
 & plants in volt-var control & 1.65 & 1.63 & 18 & trip predicted & 1 & 1 & feasible & 37.9 & 0.000 & 0 \\
 & no collector telemetry & -- & -- & 0 & insufficient data & 17 & 17 & no certifiable plant & 0.0 & 0.000 & -- \\
\bottomrule
\end{tabularx}
\end{table*}
\setcounter{table}{1}

\begin{table}[t]
\centering
\caption{Settings of the method and of the TDS validation.}
\label{tab:numerical_settings}
\scriptsize
\begin{tabularx}{\columnwidth}{>{\raggedright\arraybackslash}p{0.27\columnwidth}>{\raggedright\arraybackslash}X}
\toprule
Item & Setting \\
\midrule
Simulator & ANDES 1.9.3, implicit trapezoidal method, 2~ms step \\
Relay & Collector overvoltage, 1.10~pu, 50~ms definite time \\
Window and grid & $T=1$~s (also 2~s), 5~ms grid with intervals refined down to 0.25~ms \\
Certification margin & $\epsilon=0.05$ \\
Cascade & Eq.~\eqref{eq:cascade_iter} with propagated pickups, and trips confirmed with Eq.~\eqref{eq:dwell_excursion} \\
TDS protocol & Event at 0.5~s, simulation to 2.0~s, trips scored within 1~s of the event \\
Events & Plant trip, fixed PF ramp ($P$ and $Q$ to 70\% in 0.5~s), load disconnection, generator trip, export reduction ($\pm$20\% of the smaller unit in 1~s), reactor disconnection (60--140~MVAr), capacitor switching (55 and 85~MVAr), and line energization \\
Controls & STATCOM (100~MVAr, Kundur 50, 20~ms delay and 30~ms lag), plant reactive power command (100~ms delay), AVR setpoint ($-0.03$~pu, 50~ms delay), fast reactor (50~MVAr, 150~ms delay), and manual reactor (100~MVAr, 30~s delay) \\
Missed trip & Certified plant that trips in TDS within the window \\
\bottomrule
\end{tabularx}
\end{table}

\begin{table}[t]
\centering
\caption{Components of the 2000-bus system, counted from the RAW and DYR files and from the ANDES model as built.}
\label{tab:replica_inventory}
\scriptsize
\setlength{\tabcolsep}{3pt}
\begin{tabularx}{\columnwidth}{>{\raggedright\arraybackslash}p{0.22\columnwidth}>{\raggedright\arraybackslash}X}
\toprule
Item & Number and model \\
\midrule
Buses & 2000 in 8 areas (21 at 13.2~kV, 278 at 13.8~kV, 96 at 18~kV, 20 at 20~kV, 14 at 22~kV, 20 at 24~kV, 826 at 115~kV, 453 at 161~kV, 152 at 230~kV, and 120 at 500~kV) \\
Branches & 3206, with 2345 lines and 861 two-winding transformers \\
Synchronous machines & 435 GENROU (410 GENROU and 25 GENSAL records), of which 334 are online with 84.5~GVA \\
Exciters & 12 ESDC1A, 23 IEEET1, and 73 SEXS with default parameters replacing 61 EXPIC1, 7 ESAC6A, and 5 SCRX records. Of the 334 online machines, 84 have an exciter, and 2 exciters that cannot hold their initial field voltage are disabled \\
Governors & 367 TGOV1 replacing GGOV1 records (droop from the data), and 46 online machines without governor \\
Stabilizers & 61 IEEEST (52 on online machines), whose improper second filter is given a 1~ms double pole \\
Loads & 1350 (constant impedance in TDS), with 67.1~GW in the data and 60.4~GW after scaling area 7 \\
Shunts & 148 capacitors (17.8~GVAr) and 7 reactors (0.48~GVAr), fixed \\
Static generators & 109 without dynamic model (9.36~GW). Of the 10 in area 7, the 8 in service are represented as PVD1 plants, and the 99 outside area 7 are ideal voltage sources \\
HVDC and FACTS & None \\
Affected area & Area 7, with 431 buses, 137 online machines (24 with exciter), and a cut set of 21 branches \\
Protected plants & 16 PVD1 plants of 16--582~MW behind step-up transformers (tap 0.92087), with 2906~MW and $-967$~MVAr in total \\
Stressed state & 26 plants (5.45~GW, tap 0.92732), 12 area units online, embedded PV of 1.72~GW at 241 loads (PVD1), 1.6~GVAr of added susceptance at 230 and 500~kV buses, and loads with active power 80\% constant current and 20\% constant impedance \\
\bottomrule
\end{tabularx}
\end{table}

\subsection{Benchmark systems}\label{app:benchmarks}

Table~\ref{tab:benchmark_systems} lists the benchmark systems as simulated and Table~\ref{tab:numerical_settings} the settings of the method and of TDS. The RES share is limited by the minimum output of the synchronous units (20\% of capability or the governor minimum), whose dispatch is reduced by a common factor, and all units stay online. The step-up transformers have $x=0.12$~pu on the plant rating (1.15 times its active power) and $x/r=40$, and connect the plants to buses without generation. The plants form two to four clusters at strong buses (short-circuit ratio at least 3), and the transformer taps, within $\pm10\%$, set the collector voltages between 1.060 and 1.092~pu where this range allows. The plants' reactive power commands can increase their absorption down to power factor 0.9.

In GBnetwork, machine ratings below the active power capability are raised to the capability divided by 0.9, and the synchronous condensers are limited to an absorption of 30\% of their rating. Loads are constant power in the power flow and constant impedance in TDS, and none of the systems has OEL, UEL, or OLTC models. The linear model uses the ANDES Jacobians, completed with the derivatives of the PVD1 droop and current limit that ANDES omits (a finite-difference check of every column then finds no mismatch), and its reduction stops if $\m G_y^m$ is singular. The IEEE 39 stabilizer filters with zero time constants make the DAE of higher index, so they are given 1~ms time constants, in the method and in TDS. The NPCC data contain a slowly unstable exciter mode ($+0.011$~s$^{-1}$), which grows by 1.1\% over the 1~s window.

\subsection{2000-bus system}\label{app:activsg}

Table~\ref{tab:replica_inventory} lists the components of the 2000-bus system. The 16 plants, Granada (355~MW), Badajoz (582 and 145~MW), and 13 plants of 16--550~MW, are placed at buses of 230 and 500~kV without generation, around the bus that is electrically farthest from voltage-regulating sources. The Granada and Badajoz plants absorb the reactive power given in the report (165, 165, and 37~MVAr~\cite[pp.~14, 148, 150]{final}), the others operate at power factor 0.95 absorbing, and all of them use the PVD1 model and the transformers of the benchmarks. Besides the overvoltage stages of Section~\ref{subsec:case_activsg2000_replica}, plants have undervoltage stages of 0.75~pu for 2~s and 0.45~pu for 0.15~s~\cite{nerc2020prc024,ieee1547}, which reset when the voltage returns above $1/0.98$ times their pickup, and frequency stages at 47.5 and 51.5~Hz (50~Hz equivalent) with 0.2~s delay~\cite{eu2016rfg}. Synchronous machines have the same voltage and frequency stages and trip on pole slip.

\subsection{Protection and control changes}\label{app:ablation}

Table~\ref{tab:ablation_summary} gives the results of the method and of TDS for the changes of protection and control discussed in Section~\ref{subsec:case_intervention}.

\bibliographystyle{elsarticle-num}
\bibliography{refs}

\end{document}